\documentstyle[12pt]{article}
\def\beq{\begin{equation}}
\def\eeq{\end{equation}}
\def\bea{\begin{eqnarray}}
\def\eea{\end{eqnarray}}
\def\nn{\nonumber}
\def\ba{\begin{array}}
\def\ea{\end{array}}
\def\d{\partial}
\def\v{\vert}
\def\l{\langle}
\def\r{\rangle}
\def\one{1\hskip -1mm{\rm l}}
\def\P{{\rm l}\hskip -.85mm{\rm P}}

\begin{document}
\smallskip
\baselineskip16pt
\smallskip
\begin{center}
{\large \bf \sf
Symmetries and exact solutions \\
of some integrable Haldane-Shastry like spin chains } \\
\vspace{1.75 cm}

{\sf B. Basu-Mallick\footnote{ 
E-mail address: biru@monet.phys.s.u-tokyo.ac.jp } }

\bigskip

{\em Department of Physics, Faculty of Science, University of Tokyo, \\
 Hongo 7-3-1, Bunkyo-ku, Tokyo 113, Japan }

\end{center}

\vspace {2.5 cm}
\baselineskip=20pt
\noindent {\bf Abstract }

By using a class of `anyon like' representations of permutation algebra, 
which pick up nontrivial phase factors while interchanging the spins of two
lattice sites,  we construct some integrable variants of $SU(M)$ 
Haldane-Shastry (HS) spin chain. Lax pairs and conserved quantities 
for these spin chains are also found and it is established that these models
exhibit multi-parameter deformed or nonstandard variants of $Y(gl_M)$ 
Yangian symmetry. Moreover, by projecting the eigenstates of Dunkl operators
in a suitable way, we derive a class of exact eigenfunctions for 
such HS like spin chain and subsequently conjecture that these
exact eigenfunctions would lead to the highest weight states associated
with a multi-parameter deformed or nonstandard variant of $Y(gl_M)$ Yangian 
algebra. By using this conjecture, and acting descendent operator 
on the highest weight states associated with a nonstandard $Y(gl_2)$ Yangian
 algebra, we are able to find out the complete set of eigenvalues and 
eigenfunctions for the related HS like spin-${1\over 2}$ chain. 
It turns out that some additional energy levels, which are forbidden 
due to a selection rule in the case of $SU(2)$ HS model, interestingly 
appear in the spectrum of above mentioned  HS like spin chain having 
nonstandard $Y(gl_2)$ Yangian symmetry.
 
\newpage

\baselineskip=22pt
\noindent \section {Introduction }
\renewcommand{\theequation}{1.{\arabic{equation}}}
\setcounter{equation}{0}

\medskip

As is well known, one dimensional
integrable models with long ranged interactions have recently found wide
range of applications in diverse subjects like fractional statistics,
random matrix theory, level statistics for disordered systems,
quantum Hall effect, Yangian algebra, etc. [1-23]. In particular, the 
$SU(2)$ Haldane-Shastry (HS) spin chain [3,4], along
 with its $SU(M)$ as well as supersymmetric generalisations [15-18] 
and its deformation through twisted boundary condition [19],
provides us interesting examples of exactly solvable models which 
share many properties of an ideal gas, but with fractional statistics.
Furthermore, it is found that the conserved quantities of $SU(M)$
HS spin chain, represented  by the Hamiltonian
\beq
 H_{HS} ~=~  \sum_{ 1 \leq i <j \leq N } \,
  { 1 \over 2 \sin^2 {\pi \over N}(i-j)} \left ( P_{ij} - 1 \right ) 
\, ,
\label {a1}
\eeq
where $P_{ij}$ is the permutation operator interchanging the `spins' 
(taking $M$ possible values) of $i$-th and $j$-th lattice sites,
 lead to a realisation of $Y(gl_M)$ Yangian algebra [10].  Such $Y(gl_M)$ 
Yangian symmetry of HS spin chain (\ref {a1}) can be revealed in an elegant
  way by treating it as the `static limit' of 
dynamical spin Calogero-Sutherland (CS) model with Hamiltonian 
given by
\beq
H_{CS} ~=~  -{1\over 2} \sum_{i=1}^N ~ \left ( { \d \over \d x_i }
\right )^2 ~+
{ \pi^2 \over L^2 }~ \sum_{i<j}~{   \beta ( \beta + P_{ij}  )  \over
\sin^2 { \pi \over L} (x_i -x_j)  }~,
\label {a2}
\eeq
where $\beta $ is a coupling constant [11]. Moreover, by exploiting the 
above mentioned connection between HS spin chain and dynamical
spin CS model, it is also possible to construct a class of exact 
eigenfunctions for the spin model (\ref {a1}). It turns out that this class 
of exact eigenfunctions leads to all highest weight (HW) states of 
 $Y(gl_M)$ Yangian algebra, which appear in the reducible Hilbert space 
of HS spin chain. Consequently, by applying Yangian descendent 
operators on such HW states, one can obtain 
the complete set of eigenfunctions for the $SU(M)$ HS spin chain. 

However, it is found recently that the permutation algebra
given by
\beq
  {\cal P}_{ij}^2 ~=~ 1,
~~~{\cal P}_{ij}{\cal P}_{jl} ~=~ {\cal P}_{il}{\cal P}_{ij} ~=~
{\cal P}_{jl} {\cal P}_{il} ~, ~~~[ {\cal P}_{ij},
{\cal P}_{lm} ] ~=~ 0 \, ,
\label {a3}
\eeq
($ i, ~j,~l,~m $  being all different indices), admits a 
class of `anyon like' representations (${\tilde P}_{ij}$) which 
depend on some discrete as well as continuous parameters and
act on the spin space associated with CS model as [24] 
\beq
{\tilde P}_{ij} \,
 \v \alpha_1 \alpha_2 \cdots \alpha_i \cdots \alpha_j
\cdots \alpha_N \r ~=~ e^{ i \Phi (\alpha_i , \alpha_{i+1} , \cdots ,
\alpha_j ) } \, \v \alpha_1
\alpha_2 \cdots \alpha_j \cdots \alpha_i \cdots \alpha_N \r \,   , ~
\label {a4}
\eeq
where  $\alpha_i \, \in \, [1,2,\cdots ,M] $ and the real phase factor
$ \Phi (\alpha_i , \alpha_{i+1} , \cdots , \alpha_j ) $
is  defined in the following way. Suppose, 
$\phi_{\sigma \sigma }$s ($\sigma \, \in \, [1,2,\cdots ,M] $)
are some discrete parameters each of which may be 
chosen as $0$ or $\pi$, and 
 $~\phi_{ \gamma  \sigma }$s ($\gamma \neq \sigma $) are 
 continuous parameters which may take any real value satisfying
the antisymmetry property:
$ \phi_{ \gamma  \sigma }  =  - \phi_{ \sigma  \gamma } $. 
As a function of these $M$ number of 
discrete parameters and ${ M(M-1)/ 2 }$ number of
independent continuous parameters, the  phase factor 
 appearing in eqn.(\ref {a4}) is defined as 
\beq 
 \Phi (\alpha_i , \alpha_{i+1} , \cdots , \alpha_j )~=~
   \phi_{\alpha_i \alpha_j} ~+~ 
\sum_{\tau =1}^M  \, n_\tau \,   \left (   \phi_{\tau \alpha_j} -
 \phi_{\tau \alpha_i}    \right )  \, 
\label {a5} 
\eeq
where  $n_\tau $ ($= \sum_{p=i+1}^{j-1} \delta_{\tau \alpha_p }$)
denotes the number of occurrence of a particular spin orientation $ \tau $
in the configuration $ \v \alpha_1  \cdots \alpha_i \cdots
\alpha_p \cdots \alpha_j \cdots  \alpha_N \r $,
when the index $p$ in $\alpha_p $ runs from $i+1$ to $j-1$.
It is obvious that, the phase factor (\ref {a5}) becomes 
trivial for the simplest choice of discrete and continuous parameters 
given by:
$ \phi_{ \gamma  \gamma } ~=~ \phi_{ \gamma  \sigma }~=~0$  
for all $\sigma ,\,  \gamma $. Therefore, at this limiting case,
 ${\tilde P}_{ij}$ (\ref {a4})  would reproduce the standard 
representation of permutation algebra ($ P_{ij}$) which appears 
in the Hamiltonians (\ref {a1}) and (\ref {a2}).
However it is evident that, for all nontrivial choices of the 
parameters $ \phi_{ \gamma  \gamma }$ and $ \phi_{ \gamma  \sigma }$,  
${\tilde P}_{ij}$ (\ref {a4})  would pick up nonvanishing 
phase factors (\ref {a5}) which depend on the spin configuration 
of $(j-i+1)$ number of particles.
By using such `anyon like' representations of permutation algebra, 
 one can construct integrable variants of 
spin CS Hamiltonian  (\ref {a2}) as
\beq
{\cal H}_{CS} ~=~  -{1\over 2} \sum_{i=1}^N ~ \left ( { \d \over \d x_i }
\right )^2 ~+ { \pi^2 \over L^2 }~ \sum_{i<j}~{   
\beta ( \beta + {\tilde P}_{ij}  )  \over
\sin^2 { \pi \over L} (x_i -x_j)  }~,
\label {a6}
\eeq
 and demonstrate that this class of Hamiltonians can be solved exactly by
taking appropriate projections of the known eigenfunctions of Dunkl operators
[24]. Interestingly, it is found that the eigenvalues and eigenfunctions 
of some spin CS Hamiltonians, belonging to class (\ref {a6}),
 differ considerably from those of the standard spin CS model (\ref {a2}).
It is also revealed that, the symmetry of Hamiltonian 
(\ref {a6}) depends crucially on the discrete 
 parameters which appear in ${\tilde P}_{ij}$ (\ref {a4}), 
and may be given by an extended (i.e., multi-parameter deformed 
or nonstandard variant of) $Y(gl_M)$ Yangian algebra [25].
Moreover it turns out that the quantum $R$ matrices, 
which generate the extended $Y(gl_M)$ algebras, coincide with the 
 rational limit of some trigonometric $R$ matrices 
appearing in the algebraic Bethe ansatz 
of asymmetric six vertex model [26,27], its various generalisations [28]
and Heisenberg spin chain with Dzyaloshinsky-Moriya interaction [29].

In analogy with the case of dynamical CS models (\ref {a6}),
 we may now consider a novel class of 
HS like spin chains with the general form of Hamiltonian given by
\beq
{\cal H}_{HS} ~=~  \sum_{ 1 \leq i <j \leq N } \,
  { 1 \over {2 \sin^2 {\pi \over N}(i-j)} }
 \left ( {\tilde P}_{ij} - 1 \right ) \, ,
\label {a7}
\eeq
where ${\tilde P}_{ij}$ is an `anyon like' representation (\ref {a4}) 
of permutation algebra.  For the limiting case
${\tilde P}_{ij}= P_{ij}$, ${\cal H}_{HS}$  (\ref {a7}) evidently
reproduces the $SU(M)$ HS spin chain (\ref {a1}) 
which respects the $Y(gl_M)$ Yangian symmetry.  
So, it is natural to enquire whether the spin model (\ref {a7}) 
would also respect some Yangian like symmetry for all 
nontrivial choices of continuous and discrete parameters in 
corresponding ${\tilde P}_{ij}$. Furthermore, it should be interesting 
to investigate the nature of energy spectrum and eigenfunctions for the 
spin chain (\ref {a7}). The aim of the present article is to shed 
some light on the above mentioned issues and also reveal an intriguing 
connection between some spin chains belonging to class
(\ref {a7}) and supersymmetric versions of HS spin chain.
For this purpose, in sec.2, we start with the Lax equations for
spin chain (\ref {a7}) and use these Lax equations to  
derive the related conserved quantities. Next, in sec.3, we demonstrate that
the conserved quantities of spin chain (\ref {a7}) yield a realisation 
of extended $Y(gl_M)$ Yangian algebra. Thus it is 
established that, similar to case of dynamical CS model (\ref {a6}), 
the HS like spin chain (\ref {a7})
also respects an extended $Y(gl_M)$ Yangian symmetry. Furthermore,
by projecting the known eigenstates of Dunkl operators in a suitable way,
in sec.4 we derive a class of exact eigenvalues and eigenfunctions
for the Hamiltonian (\ref {a7}).  Subsequently, we
conjecture that the above mentioned class of exact eigenfunctions 
would lead to all HW states of extended $Y(gl_M)$ Yangian algebra, which
appear in the reducible Hilbert space of 
HS like spin chain. So, by applying appropriate descendent operators on 
such HW states one should be able to generate 
the complete set of eigenfunctions for the
 Hamiltonian (\ref {a7}). In sec.5,  we attempt to verify this 
consequence of our conjecture by applying a descendent operator on the HW
 states associated with a nonstandard $Y(gl_2)$ Yangian algebra.
In this way, we are able to find out the full spectrum
and complete set of energy eigenfunctions  for the related HS like 
spin chain with nonstandard $Y(gl_2)$  Yangian symmetry.
In sec.5, we also establish a 
relation between some spin chains, belonging to class
(\ref {a7}), and supersymmetric versions of HS spin chain.
Sec.6 is the concluding section.
\vspace{1cm}

\noindent \section { Conserved quantities of
HS like spin chains }
\renewcommand{\theequation}{2.{\arabic{equation}}}
\setcounter{equation}{0}

\medskip
In this section our aim is to find out the conserved quantities 
associated with HS like spin chain (\ref {a7}), for all 
possible choice of corresponding ${\tilde P}_{ij}$. For this purpose,
it is convenient to briefly recall the procedure [11,18] of deriving the
conserved quantities of $SU(M)$ HS spin chain (\ref {a1}) through 
the corresponding Lax pair. The operator valued elements 
of such Lax pair ($L_{HS}$ and $M_{HS}$) are given by
\beq
\left ( L_{HS} \right )_{ij} =
\left ( 1 - \delta_{ij} \right ) \theta'_{ij}  P_{ij} \, ,~~
\left ( M_{HS} \right )_{ij} = 
 -  2 \delta_{ij} \sum_{k \neq i} h'_{ik}  P_{ik}
+ 2 \left ( 1 - \delta_{ij} \right ) h'_{ij}  P_{ij} \, ,
\label {b1}
\eeq
where $ \theta'_{kj} = { \omega^k \over \omega^k - \omega^j }$,
 $h'_{kj} = \theta'_{kj} \theta'_{jk} $ and $\omega = e^{2 i \pi \over N}$.
It is easy to check that these 
operator valued elements and Hamiltonian (\ref {a1})
satisfy the following Lax equations: 
\bea
&\left [ \,  H_{HS}  \, , 
\,   \left ( L_{HS} \right )_{ij} \, \right ] ~=~ \sum_{k=1}^N
 \, \left  \{ \, \left ( L_{HS} \right )_{ik} 
\left ( M_{HS} \right )_{kj} \, - \,
 \left ( M_{HS} \right )_{ik} \left ( L_{HS} \right )_{kj}
 \right \} \, , ~~~~~~~~~~&~~(2.2a) \nn \\ 
&\left [ \,  H_{HS} \,
,  \,  X_j^{\alpha \beta } \, \right ] ~=~ \sum_{k=1}^N \,
  X_k^{\alpha \beta } \, \left ( M_{HS} \right )_{kj}  \, ,~~ 
\sum_{k=1}^N \left ( M_{HS} \right )_{jk} = 0  \, , 
~~~~~~~&(2.2b,c) \nn
\eea
\addtocounter{equation}{1}
where $X_i^{\alpha \beta }$ acts like $ \v \alpha \r \l \beta \v $
on the spin space associated with $i$-th lattice site and 
leaves the spin spaces associated with all other sites
untouched. The Lax equation (2.2c) is also called as 
`sum-to-zero' condition [8,9] for the elements of $M_{HS}$ matrix.
By using eqns.(2.2a-c), one can verify that 
the operators given by
\beq
 T_n^{ \alpha \beta } ~=~ \sum_{i,j=1}^N \, X_i^{\alpha \beta }
\left (   L_{HS}^n \right )_{ij} \, ,
\label {b3}
\eeq
where $n \in [0,1, \cdots , \infty ]$ and $\alpha , \, \beta \, \in 
[ 1, \cdots , M] $ would represent the conserved quantities of 
Hamiltonian (\ref {a1}).

Now, in close analogy with the previous case, we propose that 
the elements of Lax pair (${\cal L}_{HS}$ and ${\cal M}_{HS}$)
 associated with the HS like spin chain
(\ref {a7}) may be written as
$$
\left ( {\cal L}_{HS} \right )_{ij} =
\left ( 1 - \delta_{ij} \right ) \theta'_{ij}  {\tilde P}_{ij} \, ,~~
\left ( {\cal M}_{HS} \right )_{ij} = 
 -  2 \delta_{ij} \sum_{k \neq i} h'_{ik}  {\tilde P}_{ik}
+ 2 \left ( 1 - \delta_{ij} \right ) h'_{ij} {\tilde P}_{ij} \, , 
\eqno (2.4a,b)
$$
\addtocounter{equation}{1}
where ${\tilde P}_{ij}$ ($= \, {\tilde P}_{ji}$) is given through
eqn.(\ref {a4}). By using the permutation algebra (\ref {a3}),
it is easy to check that the Hamiltonian (\ref {a7}) 
and elements (2.4a,b) would satisfy an equation analogous to (2.2a):
\beq
\left [ \,  {\cal H}_{HS}  \, , 
\,   \left ( {\cal L}_{HS} \right )_{ij} \, \right ] ~=~ \sum_{k=1}^N
 \, \left  \{ \, \left ( {\cal L}_{HS} \right )_{ik} 
\left ( {\cal M}_{HS} \right )_{kj} \, - \,
 \left ( {\cal M}_{HS} \right )_{ik} \left ( {\cal L}_{HS} \right )_{kj}
 \right \} \, . 
\label {b5}
\eeq
Moreover, it is obvious that the elements of $ {\cal M}_{HS} $ (2.4b)
also respect the `sum-to-zero' condition: 
 $\sum_{k=1}^N \left ( {\cal M}_{HS} \right )_{jk} = 0  $.

However, one may curiously observe that an 
equation of the form: $ \left [ \,  {\cal H}_{HS} 
,    X_j^{\alpha \beta } \, \right ] \, = \, \sum_{k=1}^N \,
  X_k^{\alpha \beta } \, \left ( {\cal M}_{HS} \right )_{kj}  \,$,
is no longer satisfied except for the limiting case 
${\tilde P}_{ij}=  P_{ij}$. So, it is necessary
to find out an appropriate generalisation
of operator $ X_i^{\alpha \beta } $ which would satisfy a Lax equation
analogous to (2.2b).  To this end, 
we notice that the `anyon like' representation (\ref {a4}) can also 
be expressed through the operator relation 
\beq
{\tilde P}_{i,i+1}  =  Q_{i,i+1}P_{i,i+1}  , ~
{\tilde P}_{il}  =  S_{il} {\tilde P}_{l-1,l}S_{il}^{-1} = 
\left( Q_{i,i+1}  Q_{i,i+2}  \cdots  Q_{il} \right)   P_{il} 
\left(  Q_{l-1,i} Q_{l-2,i}  \cdots   Q_{i+1,i}  \right) ,
\label {b6}
\eeq
where $l >i+1$, $ S_{il} = {\tilde P}_{i,i+1} {\tilde P}_{i+1,i+2} \cdots  
{\tilde P}_{l-2,l-1} $ and  $Q_{ik}$, which acts like a matrix $Q$
 on the direct product of
 $i$-th and $k$-th spin spaces (but acts trivially on all
other spin spaces), is given by
\beq
Q_{ik}~=~ \sum_{\sigma =1}^M \, e^{i\phi_{\sigma \sigma }}
 \, X_i^{ \sigma \sigma }\otimes X_{k}^{ \sigma \sigma }~+~
\sum _{ \sigma \neq \gamma } \, e^{i\phi_{\gamma \sigma }}
 \, X_i^{ \sigma \sigma }\otimes X_{k}^{ \gamma \gamma } \, .
\label {b7}
\eeq
It should be noted that the above $Q$ matrix contains all 
continuous as well as discrete parameters which are present 
in eqn.(\ref {a4}). So there exists a unique correspondence 
between this $Q$ matrix and `anyon like' representation 
(\ref {a4}) of permutation algebra.
At the limiting case $Q_{ik} = \one $, which is 
obtained from eqn.(\ref {b7}) by choosing 
$ \phi_{ \gamma  \gamma } ~=~ \phi_{ \gamma  \sigma }~=~0$  
for all $\sigma ,\,  \gamma $,
${\tilde P}_{ij}$ (\ref {b6}) evidently reproduces the standard 
representation $P_{ij}$. Moreover it can be checked that, 
for any possible choice of related continuous and discrete
parameters, the $Q$ matrix (\ref {b7}) satisfies the following 
two equations:
\beq
Q_{ik} \, Q_{il} \, Q_{kl} ~=~ Q_{kl} \, Q_{il} \, Q_{ik}
\, , ~~~~~ Q_{ik} \, Q_{ki}  ~=~ \one \, .
\label {b8}
\eeq
With the help of these two equations, one can
interestingly prove that ${\tilde P}_{ij}$ (\ref {b6}) indeed gives a 
valid representation of permutation algebra (\ref {a3}). 

By using the $Q$ matrix (\ref {b7}),
we may now define a set of nonlocal operators as
\beq
 {\tilde X}_i^{ \alpha \beta } ~=~ Q_{i,i+1} Q_{i,i+2} \cdots
 Q_{iN}  \, X_i^{\alpha \beta } \,
  Q_{i1} Q_{i2} \cdots Q_{i,i-1} \, ,
\label {b9}
\eeq
which would reduce to $X_i^{\alpha \beta }$
 at the limiting case $Q= \one $ and, therefore, might be considered
as some generalisation of $X_i^{\alpha \beta }$.
With the help of eqn.(\ref {b8}),
it can be shown that ${\tilde X}_i^{ \alpha \beta }$ (\ref {b9})
and ${\tilde P}_{ij}$ (\ref {b6}) obey the following 
simple relation:
\beq
 {\tilde P}_{ij}{\tilde X}_i^{ \alpha \beta } ~=~ 
 {\tilde X}_j^{ \alpha \beta } {\tilde P}_{ij} \, .
\label {b10}
\eeq
Furthermore, by applying the above relation, it is easy to verify that 
the Hamiltonian  (\ref {a7}), $ {\tilde X}_i^{ \alpha \beta } $ 
(\ref {b9}) and $\left ( {\cal M}_{HS} \right )_{ij}$ (2.4b)
satisfy an equation given by
\beq
\left [ \,  {\cal H}_{HS} \,
,  \,  {\tilde X}_j^{\alpha \beta } \, \right ] ~=~ \sum_{k=1}^N \,
  {\tilde X}_k^{\alpha \beta } \, \left ( {\cal M}_{HS} \right )_{kj}  \, .
\label {b11}
\eeq
So we curiously find that, through the introduction of
 nonlocal operator $ {\tilde X}_i^{\alpha \beta }$, 
it is possible to obtain the above Lax  
equation which is an analogue of (2.2b) for the present case. 
Now, by applying 
 Lax equations (\ref {b5}) and (\ref {b11}), and also using 
 `sum-to-zero' condition satisfied by the
 elements of $ {\cal M}_{HS} $ (2.4b), one can verify that
the operators given by
\beq
 {\cal T}_n^{ \alpha \beta } ~=~ \sum_{i,j=1}^N \, 
{\tilde X}_i^{\alpha \beta }
\left (   {\cal L}_{HS}^n \right )_{ij} \, ,
\label {b12}
\eeq
where $n \in [0,1, \cdots , \infty ]$ and $\alpha , \, \beta \, \in 
[ 1, \cdots , M] $, would commute with the Hamiltonian (\ref {a7})
of HS like spin chain.  At the limiting case $Q= \one $, 
 ${\cal T}_n^{ \alpha \beta } $ (\ref {b12}) evidently reproduces 
the local conserved quantities  
(\ref {b3}) of $SU(M)$ HS spin chain (\ref {a1}). 

Thus, we are able to derive here the Lax equations and conserved quantities 
associated with spin chains belonging to the class
(\ref {a7}). The $Q$ matrix (\ref {b7}) and  related 
nonlocal operators ${\tilde X}_i^{\alpha \beta }$ (\ref {b9})
play a crucial role in our derivation. 
In the next section, our aim is 
to find out the symmetry algebra generated by the conserved 
quantities (\ref {b12}) and establish their connection with the
Yang-Baxter equation.

\vspace{1cm}

\noindent \section { Symmetries of HS like spin chains }
\renewcommand{\theequation}{3.{\arabic{equation}}}
\setcounter{equation}{0}

\medskip

It is already mentioned that the symmetry of $SU(M)$ HS spin chain 
is given by the $Y(gl_M)$ Yangian algebra. 
This  $Y(gl_M)$ Yangian algebra [30,31]
may be defined through the operator valued elements of an
$M\times M$ dimensional monodromy matrix $T^0(u)$,
which obeys the quantum Yang-Baxter equation (QYBE)
\beq
R_{00'}(u-v) \left ( T^0(u) \otimes \one \right )
 \left ( \one \otimes  T^{0'}(v) \right ) ~=~
 \left ( \one \otimes  T^{0'}(v) \right )
 \left ( T^0(u) \otimes \one \right )  R_{00'}(u-v) \, .
\label {c1}
\eeq
Here $u$ and $v$ are  spectral parameters and
the $M^2 \times M^2$ dimensional rational $R(u-v)$ matrix,  having 
$c$-number valued elements, is taken as
\beq
R_{00'} (u-v)  ~=~   (u-v) \, \one  \, + \, P_{00'} \, .
\label {c2}
\eeq
Associativity of $Y(gl_M)$
algebra is ensured from the fact that the above $R$ matrix 
satisfies the Yang-Baxter equation (YBE) given by
\beq
  R_{00'} (u-v) \, R_{00''} (u-w) \, R_{0'0''} (v-w)  ~=~
  R_{0'0''} (v-w) \, R_{00''} (u-w) \, R_{00'} (u-v) \,  ,
\label {c3}
\eeq
where a matrix like $R_{ab}(u-v)$ acts nontrivially only on the $a$-th
and $b$-th vector spaces. The conserved quantities (\ref {b3}) of 
$SU(M)$ HS spin chain yield a realisation of the above
mentioned monodromy matrix which generates the $Y(gl_M)$ Yangian algebra.

Now, for finding out the symmetry algebra associated with 
conserved quantities (\ref {b12}),
we consider a  rational $R$ matrix of the form
$$
R_{00'} (u-v)  ~=~   (u-v) \, Q_{00'} \, + \,  P_{00'} \, ,
 \eqno (3.4a)  
$$
where $Q_{00'}$ acts exactly like $Q_{ik}$ (\ref {b7})
on the direct product of $0$-th and $0'$-th auxiliary spaces: 
$$
Q_{00'}~=~ \sum_{\sigma =1}^M \, e^{i\phi_{\sigma \sigma }}
 \, X_0^{ \sigma \sigma }\otimes X_{0'}^{ \sigma \sigma }~+~
\sum _{ \sigma \neq \gamma } \, e^{i\phi_{\gamma \sigma }}
 \, X_0^{ \sigma \sigma }\otimes X_{0'}^{ \gamma \gamma } \, .
\eqno (3.4b) 
$$
\addtocounter{equation}{1}
It is interesting to notice that, the above defined  
$R$ matrix (3.4a) coincides with the rational limit 
of a trigonometric solution of YBE related to some 
asymmetric vertex models [28].
Moreover, by using the conditions (\ref {b8}) satisfied by $Q$ matrix, 
it is easy to directly check that the $R$ matrix (3.4a) would be a valid
  solution of YBE (\ref {c3}).  
At the limiting case $Q_{00'} = \one $, the rational solution 
(3.4a) obviously reduces to $R$ matrix (\ref {c2}) which 
leads to $Y(gl_M)$ Yangian algebra through QYBE. So it is clear
that by substituting the $R$ matrix (3.4a) to QYBE (\ref {c1}), 
and expressing this QYBE through the elements of related monodromy
matrix, one can generate an associative Yangian like algebra.
All discrete and continuous 
parameters, which are present in the $Q$ matrix (3.4b), would naturally
appear in the defining relations of above mentioned `extended' $Y(gl_M)$ 
Yangian algebra. Depending on the choices of such discrete parameters 
($\phi_{\sigma \sigma }$),
one can also make a  rather useful classification
of corresponding extended Yangian algebras. For this purpose it may be 
noted that, if the discrete parameters are chosen as 
$\phi_{\sigma \sigma }= 0$ (or, $\phi_{\sigma \sigma }= \pi$) 
for all $\sigma$, the $Q$ matrix (3.4b) can be rewritten in the form 
\beq
Q_{00'}~=~ \epsilon \, \one ~+~ 
\sum _{ \sigma \neq \gamma } \, \left ( \,  e^{i\phi_{\gamma \sigma }}
- \epsilon \right )\, 
X_0^{ \sigma \sigma }\otimes X_{0'}^{ \gamma \gamma }\, ,
\label {c5}
\eeq
where $\epsilon = 1$ (or, $\epsilon = -1$). 
Evidently the extended Yangian algebra, generated by the $Q$ matrix 
(\ref {c5}) or corresponding $R$ matrix (3.4a), would reduce to the standard 
$Y(gl_M)$ algebra at the limit 
$\phi_{\gamma \sigma } \rightarrow 0$ 
(or, $ \phi_{\gamma \sigma } \rightarrow \pi $)
of all continuous parameters. So, it is natural to classify 
 these extended algebras as `multi-parameter dependent deformations' 
of $Y(gl_M)$ Yangian algebra. A few concrete examples of this type of
 multi-parameter deformed Yangian algebras have already 
appeared in the literature [32-34] and it is recently found that
an integrable extension of Hubbard model respects such deformed Yangian 
symmetry [35].  However it may be easily checked that, if all
discrete parameters appearing in the $Q$ matrix (3.4b) 
do not take the same value,
the corresponding extended Yangian algebras would not reduce to 
standard $Y(gl_M)$ algebra at any limit of the related 
continuous parameters. So we may classify such extended algebras 
as `nonstandard' variants of $Y(gl_M)$ Yangian algebra.

At present, 
we want to show that the symmetry of HS like spin chain 
(\ref {a7}) would be given by the extended $Y(gl_M)$ Yangian algebra 
 associated with $R$ matrix (3.4a). To this end, we start with
 the Hilbert space ($S$) of HS like spin chain 
(\ref {a7}) where a general state vector is written as
 \beq
\v \Psi \r ~=~ \sum_{ \{ \alpha_i \}  } ~
 \Psi_{ \alpha_1  \alpha_2 \cdots  \alpha_N  } 
\, \v \alpha_1 \alpha_2 \cdots \alpha_N \r \, ,
\label {c6}
\eeq
and, by following Ref.11, define another space 
$\Gamma $ where a general state vector is given by
\beq
\v \chi \r ~=~ \sum_{ \{ \alpha_i \} , \{ n_i \} } ~
 \chi^{ \alpha_1  \alpha_2  \cdots  \alpha_N  }_{n_1 n_2  \cdots  n_N}
\, \v \alpha_1 \alpha_2 \cdots \alpha_N \r \otimes 
\, \v n_1 n_2 \cdots n_N \r  \, 
\label {c7}
\eeq
with $\alpha_i \in [1,2,\cdots ,M]$  and 
$ n_1 \, n_2  \, \cdots \, n_N $  being a permutation of 
$1\, 2 \, \cdots \, N$.
So, the space $\Gamma $ is obtained by enlarging the Hilbert space $S$ through 
some discrete coordinate ($n_i$) degrees of freedom. One may 
now define some operators which act nontrivially only on the
coordinate part of state vectors associated with space $\Gamma $:
\bea
&{\hat z}_i \, \v n_1  n_2  \cdots  n_i  \cdots n_N \r ~=~\omega^{n_i}\,
 \v n_1  n_2  \cdots  n_i  \cdots n_N \r  \, , ~~\nn \\
&K_{ij} \,  \v n_1  n_2  \cdots  n_i \cdots n_j  \cdots n_N \r 
~=~ \v n_1  n_2  \cdots  n_j \cdots n_i  \cdots n_N \r  \, ,
\label {c8}
\eea
where $\omega= e^{2 i \pi \over N}$.
It is easy to check that the above defined operators 
satisfy the algebraic relations given by
\bea
&~~  K_{ij}{\hat z}_i ~=~ 
{\hat z}_j K_{ij} \, ,~~K_{ij} {\hat z}_l ~=~ {\hat z}_l K_{ij} \, , 
\nn   \\
&~~ K_{ij}^2 ~=~ \one  \, , ~~~K_{ij}K_{jl} ~=~ K_{il}K_{ij} ~=~
K_{jl} K_{il} \, , ~~~[\, K_{ij}, K_{lm} \, ] ~=~ 0 \,,~ 
\label {c9}
\eea
$ i, ~j,~l,~m ~$ being all different indices. With the help of 
these  ${\hat z}_i ,~K_{ij},$ one can define Dunkl operators  of the form
\beq
 D_i ~=~ \sum_{j \ne i } \, {\hat \theta}_{ij} \,  K_{ij}  ~ ,
\label {c10}
\eeq
where $ {\hat \theta}_{ij} = { {\hat z}_i 
\over {\hat z}_i - {\hat z}_j } $ and 
show that such Dunkl operators satisfy the 
commutation relations given by
\beq
K_{ij} D_i ~= ~ D_j K_{ij} \, , ~~\left [ \, K_{ij} , D_k \,
 \right ] ~=~ 0 \, , ~~\left [ \, D_i , D_j \, \right ] ~=~
 \left ( D_i -D_j \right ) K_{ij} \, ,
\label {c11}
\eeq
where $k\neq i,j $. These commutation relations will be used shortly
for finding out a concrete 
realisation of extended $Y(gl_M)$ algebra.

Next, we consider all state vectors of $\Gamma $ which satisfy the 
following relation: 
\beq
 \left ( K_{ij} - {\tilde P}_{ij} \right ) \v \chi \r ~=~ 0 \, ,
\label {c12}
\eeq
where the action of ${\tilde P}_{ij}$ is defined by eqn.(\ref {a4}).
 These state vectors would form a subspace of $\Gamma $, which we
denote by $\Gamma_B^* $. By using eqns.(\ref {c12}) and (\ref {c7}),
it is easy to verify that coefficients of $ \v \chi \r $ now satisfy 
an `anyonic' symmetry condition:
\beq
\chi^{ \alpha_1  \alpha_2  \cdots \alpha_i \cdots \alpha_j \cdots 
\alpha_N }_{n_1 n_2  \cdots n_i \cdots n_j  \cdots  n_N} ~=~
e^{-i \Phi \left ( \alpha_i , \alpha_{i+1} , \cdots , \alpha_j \right )}
 \, \chi^{ \alpha_1  \alpha_2  \cdots \alpha_j \cdots \alpha_i \cdots 
\alpha_N }_{n_1 n_2  \cdots n_j \cdots n_i  \cdots  n_N} \, ,
\label {c13}
\eeq
where $ \Phi \left ( \alpha_i , \alpha_{i+1} , \cdots , \alpha_j \right )$
is given by (\ref {a5}). Due to the existence of above symmetry 
condition, only the knowledge of  coefficients like 
 $\chi^{ \alpha_1  \alpha_2  \cdots  \alpha_N  }_{1 2 \cdots \cdots  N} $ 
is sufficient to construct any 
 $ \v \chi \r $ belonging to space $\Gamma_B^* $.
Consequently, it is possible to establish an isomorphism 
between the spaces  $\Gamma_B^*$ and $S$ through the 
 one-to-one mapping: $\pi^* \v \chi \r \, = \, \v \Psi \r \, $, where 
the coefficients of $ \v \Psi \r $ and $ \v \chi \r $ are related as 
\beq 
 \Psi_{ \alpha_1  \alpha_2 \cdots  \alpha_N  } ~=~
 \chi^{ \alpha_1  \alpha_2  \cdots  \alpha_N  }_{1 2 
\cdots \cdots  N} \, .
 \label {c14}
\eeq
It is clear from eqn.(\ref {c13}) that, at the limiting case $\Phi = 0$, 
$\Gamma_B^* $  would represent a bosonic space ($\Gamma_B$) 
where the wave functions remain invariant under the simultaneous interchange 
of spin and coordinate degrees of freedom.
 Thus, at this limiting case,  $\pi^* $ would
reproduce a mapping $\pi $ which was used earlier [11] 
for defining an isomorphism between the spaces $\Gamma_B$ and $S$.

Now, by using the Dunkl operators (\ref {c10})
and $Q$ matrix (3.4b), we propose the following monodromy matrix 
whose elements are some operators on space $\Gamma_B^*$:
\beq
{\hat T}^0(u) ~=~  \Omega_0 +  \sum_{i=1}^N {\P_{0i} \over u- D_i } \, ,
\label {c15}
\eeq
where 
$$
~\Omega_0 \, = \, Q_{01} Q_{02} \cdots Q_{0N} \, , ~~
\P_{0i} \,  =  \, \left ( Q_{01} Q_{02} \cdots Q_{0,i-1} \right ) \, P_{0i} \,
\left ( Q_{0,i+1} Q_{0,i+2} \cdots Q_{0N} \right )   \, .
\eqno (3.16a,b)
$$
\addtocounter{equation}{1}
With the help of eqns.(\ref {b8}) and 
 (\ref {b6}), it may be checked that: $ \left [ \, {\hat T}^0(u) \, , \,
K_{ij} {\tilde P}_{ij} \,  \right ] = 0 $. Consequently,
 the monodromy matrix ${\hat T}^0(u) $  (\ref {c15}) 
preserves the `anyonic' symmetry (\ref {c13}) of state vectors belonging to
 space $\Gamma_B^* $. So, by using the mapping 
 $\pi^* $ defined through eqn.(\ref {c14}), 
we may construct an image ($ \pi^* [ {\hat T}^0(u)]$) 
 of the monodromy matrix (\ref {c15}) on space $S$ as
\beq
T^0(u) ~ \equiv ~ \pi^* \left [ {\hat T}^0(u) \right ] 
~=~ \pi^* \,
\left( \Omega_0 +  \sum_{i=1}^N {\P_{0i} \over u- D_i }\right )
 \, \left ( \pi^* \right )^{-1} \, .
\label {c17}
\eeq
It is worth noting that, at the limiting 
case $Q = \one $, eqn.(\ref {c17}) reduces to
$$
 T^0(u) 
~=~ \pi \, \left( \one + \sum_{i=1}^N { P_{0i} \over u- D_i }\right )
  \, \pi^{-1} \,,
$$
which represents the monodromy matrix of $SU(M)$ HS spin chain 
(\ref {a1}) and yields a solution of
QYBE (\ref {c1}) when the corresponding $R$ matrix is taken as (\ref {c2})  
[11].  Now, by  essentially following the approach of Ref.11 and using 
the relations (\ref {c11}) as well as (\ref {b8}), one can 
interestingly show that the monodromy matrix (\ref {c17}) would also satisfy 
QYBE (\ref {c1})
when the corresponding $R$ matrix is taken as (3.4a).
Thus,  it turns out that the monodromy matrix (\ref {c17}) yields 
a concrete realisation of extended $Y(gl_M)$ Yangian algebra.

Next, we want to derive a few  simple relations which are needed 
 for establishing a connection between the conserved quantities (\ref {b12})
and our realisation (\ref {c17}) of extended Yangian algebra.
First of all, by using the standard
expression: $ P_{0i} = \sum_{\alpha , \beta = 1}^M \, 
X_0^{\alpha \beta } $   $ \otimes  X_i^{\beta \alpha } \,$
and conditions (\ref {b8}) on $Q$ matrix, we find that the operator
$\P_{0i}$ (3.16b)  may be rewritten in a compact form given by
\beq
\P_{0i} ~=~ \sum_{\alpha , \beta = 1}^M \,
X_0^{\alpha \beta } \otimes {\tilde X}_i^{\beta \alpha } \, .
\label {c18}
\eeq
With the help of above expression and commutation relations (\ref {c11}),
one can further show that
\beq
 \pi^*  \left [ \, \sum_{i=1}^N  \P_{0i}   D_i^n \, \right ] ~=~
 \pi^* \, \left ( \, \sum_{i=1}^N  \P_{0i}   D_i^n \, \right ) \,
  \left ( \pi^* \right )^{-1} ~=~
X_0^{\alpha \beta } \otimes {\cal T}_n^{\beta \alpha } \, ,
\label {c19}
\eeq
where 
$ {\cal T}_n^{\beta \alpha } $s are the conserved quantities (\ref {b12})
of HS like spin chain. Finally, by using eqn.(\ref {c19}), 
we find that the monodromy matrix (\ref {c17}) can be expressed 
through its modes as
\beq
 T^0(u) ~=~ \Omega_0 \, + \,  \sum_{n=0}^\infty  \,
{1\over u^{n+1} } \, \sum_{\alpha ,\beta = 1}^M  \,
\left (\, X_0^{\alpha \beta } \otimes
 {\cal T}_n^{\beta \alpha } \, \right )  ~ .
\label {c20}
\eeq
One may curiously notice that, 
apart from the conserved quantities (\ref {b12}) which 
are derived from the Lax equations, 
 the operator $\Omega_0$  also appears in the mode expansion 
(\ref {c20}).  But it seems that, by
 applying  Lax equations ((\ref {b5}),(\ref {b11})), it is not possible
 to determine whether this $\Omega_0$  would also commute 
with the spin chain Hamiltonian (\ref {a7}).
However, with the help of eqns.(3.16a), (\ref {b6}) and 
(\ref {b8}), one can directly show that
\beq
 \left [ \, \Omega_0  \, , \, {\tilde P}_{ij} \, \right ] ~=~ 
 \left [ \, \Omega^{\alpha  \alpha }
  \, , \, {\tilde P}_{ij} \, \right ] ~=~ 0 \, ,  
\label {c21}
\eeq
where 
 $ \Omega^{ \alpha \alpha }$s  are operator valued elements of the 
diagonal matrix $\, \Omega_0$: $~ \Omega_0 \, 
= \sum_{\alpha = 1 }^M \,  X_0^{\alpha \alpha } \otimes 
\Omega^{\alpha \alpha } \, $. By using the 
 relation (\ref {c21}), we may now easily verify that  
$\Omega_0$  and $ \Omega^{ \alpha \alpha }$
indeed commute with the Hamiltonian (\ref {a7}).  Thus  
we observe that,  HS like spin chains apparently possess 
some additional conserved quantities ($\Omega^{\alpha \alpha }$) which
become trivial at $Q= \one$ limit, i.e., for the case of $SU(M)$ HS model.

From the above discussion it is revealed that, all modes
of the monodromy matrix $ T^0(u)$ (\ref {c20}) 
can be identified with various conserved 
quantities which commute with the Hamiltonian (\ref {a7}). 
Therefore, we may conclude that the HS like spin chain (\ref {a7})
respects an extended $Y(gl_M)$ Yangian symmetry. 
Furthermore, one can explicitly find out all algebraic relations among the
conserved quantities of such spin chain, 
by simply expressing this extended $Y(gl_M)$ Yangian algebra
through the modes of corresponding monodromy matrix. To this end, 
we insert the mode expansion (\ref {c20}) as well as 
 $R$ matrix (3.4a) to QYBE (\ref {c1}), and, equating from its both 
sides the coefficients of same  powers in  $u,~v$,  easily 
obtain the following set of relations:
\bea
     &\left[\, \Omega^{\alpha \alpha } \, , \, \Omega^{\beta \beta } \, 
\right ] ~=~0~,~~~~ \Omega^{\alpha \alpha } \, 
{\cal T}_n^{ \beta \gamma } ~=~ 
{\rho_{ \alpha \gamma } \over \rho_{\alpha \beta }} \,
{\cal T}_n^{ \beta \gamma } \Omega^{\alpha \alpha } \, , ~~
\nn &(3.22a,b) ~~~~~~
\\
 &\rho_{ \alpha \gamma } \, {\cal T}^{\alpha \beta }_0 \, 
{\cal T}^{ \gamma \delta }_n 
~-~\rho_{\beta \delta } \, {\cal T}^{\gamma \delta }_n  \, 
{\cal T}^{\alpha \beta }_0 
~~=~~
\delta_{ \alpha \delta } \, {\cal T}^{\gamma \beta }_n \, 
\Omega^{\alpha \alpha  }~-~\delta_{ \gamma \beta }
~\Omega^{ \gamma \gamma } \, {\cal T}^{\alpha \delta }_n \,,~ 
\nn  &~~(3.22c)~~~~~~
 \\
 &\left [ ~\rho_{ \alpha \gamma } \, {\cal T}^{ \alpha \beta }_{n+1} \, 
 {\cal T}^{\gamma \delta }_m  \, - \, \rho_{\beta \delta } \,
{\cal T}^{ \gamma \delta }_m \, {\cal T}^{ \alpha \beta }_{n+1}~ \right ] ~-~
 \left [ ~\rho_{ \alpha \gamma } \, {\cal T}^{ \alpha \beta }_{n} \, 
 {\cal T}^{\gamma \delta }_{m+1}  \, - \, \rho_{\beta \delta } \,
{\cal T}^{ \gamma \delta }_{m+1} \, {\cal T}^{ \alpha \beta }_n~ \right ] \nn
&\\ &\hskip 8 cm ~=~   ~\left (~ {\cal T}^{ \gamma \beta }_m \,
{\cal T}^{ \alpha \delta }_n  ~-~  
{\cal T}^{ \gamma \beta }_n \, {\cal T}^{ \alpha \delta }_m~
\right ) \, , \nn &~~(3.22d) ~~~~~~
\eea
\addtocounter{equation}{1}
where $\rho_{ \alpha \gamma } = e^{ i\phi_{ \alpha \gamma }} $. Thus 
we are able to explicitly derive here the symmetry algebra associated with
 HS like spin chain (\ref {a7}).
It may be noted that, at the special case
 $\rho_{ \alpha \gamma } = 1 $ and $\Omega^{\alpha \alpha } = \one $ 
for all $\alpha , ~\gamma $, the algebra (3.22)
reduces to standard $Y(gl_M)$ Yangian and reproduces the 
commutation relations among the conserved quantities of $SU(M)$
 HS spin chain.

\vspace{1cm}

\noindent \section { A class of exact eigenfunctions for HS like spin chains }
\renewcommand{\theequation}{4.{\arabic{equation}}}
\setcounter{equation}{0}

\medskip

As was mentioned in sec.1, by exploiting a 
connection between spin chain (\ref {a1}) and dynamical CS model 
(\ref {a2}), one can construct a class of exact eigenfunctions for the 
$SU(M)$ HS spin chain  [11]. 
In the following, our aim is to pursue a similar approach for
 the HS like spin chain (\ref {a7}) and construct its 
eigenfunctions which may be related 
to the HW states of extended $Y(gl_M)$ algebra. To this end, we
consider again the spin space $S$  where a general state vector is 
written in the form (\ref {c6}). Moreover, we fix our convention by saying 
that the  $i$-th lattice site contain `up' spin if $\alpha_i = 1$, and 
`down' spin if $\alpha_i \neq 1$. Notice that any state vector of $S$,
containing only $m$ number of `down' spins,
may also be expressed in the form:
$
\v \Psi_{m} \r ~=~ \sum_{ \{ \gamma_i \} , \{ n_i \} } ~
 \psi^{ \gamma_1  \gamma_2  \cdots  \gamma_m  }_{n_1 n_2  \cdots  n_m}
\, \v \gamma_1 \gamma_2 \cdots \gamma_m \r \otimes 
\, \v n_1 n_2 \cdots n_m \r  \,  ,
$
where $n_i$ denotes the position of $i$-th `down' spin 
($ 1\leq n_1 < n_2 < \cdots < n_m \leq N $) and 
$\gamma_i \in [2,3,\cdots ,M]$ represents the precise orientation of 
this `down' spin.  Such $m$-magnon states will span a 
subspace  of $S$, which we denote by $S_m$.  

Next, we 
define another space ${\cal S}_m$, where the form of a general 
state vector is apparently same as the above described 
$m$-magnon state:
\beq
\v \psi_{m} \r ~=~ \sum_{ \{ \gamma_i \} , \{ n_i \} } ~
 \psi^{ \gamma_1  \gamma_2  \cdots  \gamma_m  }_{n_1 n_2  \cdots  n_m}
\, \v \gamma_1 \gamma_2 \cdots \gamma_m \r \otimes 
 \v n_1 n_2 \cdots n_m \r  \,  .
\label {d1}
\eeq
However, at present, there exist no ordering among
 $n_i$s which appear in $\v n_1 n_2 \cdots $   $ n_m \r $, and  these
$n_i$s may freely take all possible distinct values 
(i.e., $ n_i \neq n_j $ when $i \neq j$ ) ranging from $1$ to $N$. 
Moreover, the coefficients like 
$ \psi^{ \gamma_1  \gamma_2  \cdots  \gamma_m  }_{n_1 n_2  \cdots  n_m}$
are required to satisfy the following `anyonic' symmetry condition: 
\beq
\psi^{ \gamma_1  \gamma_2  \cdots \gamma_i \cdots \gamma_j \cdots 
\gamma_m }_{n_1 n_2  \cdots n_i \cdots n_j  \cdots  n_m} ~=~
e^{-i \Phi \left ( \gamma_i , \gamma_{i+1} , \cdots , \gamma_j \right ) }
 \, \psi^{ \gamma_1  \gamma_2  \cdots \gamma_j \cdots \gamma_i \cdots 
\gamma_m }_{n_1 n_2  \cdots n_j \cdots n_i  \cdots  n_m} \, ,
\label {d2}
\eeq
where $ \Phi \left ( \gamma_i , \gamma_{i+1} , \cdots , \gamma_j \right )$
is obtained from (\ref {a5}) by restricting 
$\alpha_i \in [2,3,\cdots ,M]$. So the condition 
$ \left ( K_{ij} - {\tilde P}_{ij} \right ) \v \psi_m \r = 0 $ is satisfied
for all state vectors of ${\cal S}_m$.  It is clear from eqn.(\ref {d2})
that, only the knowledge of  coefficients
 $\psi^{ \gamma_1  \gamma_2  \cdots  \gamma_m  }_{n_1 n_2  \cdots  n_m}$,
with $ 1\leq n_1 < n_2 < \cdots < n_m \leq N $, is sufficient to 
construct the state vector $\v \psi_{m} \r $
 (\ref {d1}). Consequently, there exists an isomorphism between the 
spaces $S_m$ and ${\cal S}_m$,  and one might regard
${\cal S}_m$ itself as the $m$-magnon space. So, we may consider a 
mapping $p_m$, which projects the state vectors of $S$ to
the $m$-magnon space ${\cal S}_m$: $p_m \v \Psi \r = \v \psi_m \r $, 
where the coefficients of $\v \psi_{m} \r $ are defined for the case 
$ 1\leq n_1 < n_2 < \cdots < n_m \leq N $ as
\beq 
\psi^{ \gamma_1  \gamma_2  \cdots  \gamma_m  }_{n_1 n_2  \cdots  n_m}
~=~ \Psi_{ \alpha_1  \alpha_2 \cdots  \alpha_N  } \, ,
\label {d3}
\eeq
with $ \, \alpha_{n_1} = \gamma_1 , ~ \alpha_{n_2} = \gamma_2  , \cdots ,
\alpha_{n_m} = \gamma_m $ and $\alpha_i = 1 $ if $i \neq n_1, n_2 , 
\cdots ,  n_m $. The mapping $p_m$ can also be used to project
the Hamiltonian (\ref {a7}) of HS like spin chain on space 
${\cal S}_m$. Since the Hamiltonian (\ref {a7}) preserves the 
$m$-magnon space, its projection is unambiguously
 defined through the relation: 
\beq
p_m \left [ {\cal H}_{HS} \right ] \cdot p_m \v \Psi \r ~=~ 
p_m \left ( \, {\cal H}_{HS} \v \Psi \r \,  \right ) \, ,
\label  {d4}
\eeq
where $\v \Psi \r $ is an arbitrary state vector of space $S$. 
Thus, the problem 
of diagonalising Hamiltonian (\ref {a7}) now reduces to the problem of 
diagonalising  $ p_m \left [ {\cal H}_{HS} \right ] $ in all 
$m$-magnon sectors.

Though it is possible, in principle, to explicitly
find out the matrix elements of 
$ p_m \left [ {\cal H}_{HS} \right ] $ 
 by directly using eqn.(\ref {d4}), we wish to 
 apply an easier alternative method for this purpose.
To this end, however, we find that it is 
convenient to fix a few continuous 
and discrete parameters appearing in the phase factor (\ref {a5}) as 
\beq
\phi_{11}~ =~ \phi_{1\gamma }= ~\phi_{\gamma 1}~ =~ 0 \, ,
\label {d5}
\eeq
 where $\gamma \in [2,3, \cdots M]$. The above choice of 
 continuous and discrete parameters evidently gives a special 
status to the `up' spin component  and implies that the phase factor 
(\ref {a5}) will vanish while interchanging this 
`up' spin with any spin component (`up' or `down') sitting at the 
nearest lattice site. It should be noted that, even after fixing 
a few discrete and continuous parameters through eqn.(\ref{d5}),
the phase factor (\ref {a5}) and related `anyon like' representation 
(\ref {a4}) still contain $(M-1)$ number of free 
discrete parameters and $(M-1)(M-2)/2$ number of free continuous 
parameters.  Our discussion in the following
would be valid for all possible choices of these remaining 
discrete as well as continuous parameters in the Hamiltonian (\ref {a7}).

Next, we consider another mapping $q_m$ which projects the state vectors 
of $\Gamma_B^*$ to space ${\cal S}_m$: 
$q_m  \, \v \chi \r \, = \, \v \psi_m \r $, where the coefficients of  
$\v \psi_m \r $ (\ref {d1}) and $\v \chi \r  $ (\ref {c7}) are related as 
\beq
\psi^{ \gamma_1  \gamma_2  \cdots  \gamma_m  }_{n_1 n_2  \cdots  n_m}~=~
\chi^{ \gamma_1  \gamma_2  \cdots  \gamma_m  \, 1 \, 1  \,
\cdots \cdots  \cdots \cdots \cdots \, 1 }_{
n_1 n_2  \cdots  n_m n_{m+1} n_{m+2} \cdots n_N } \, ,
\label {d6}
\eeq
with $n_1 n_2  \cdots  n_m n_{m+1} n_{m+2} \cdots n_N $ being a permutation
of $1 \, 2 \, \cdots \, N$. Due to eqns.(\ref {c13}) and (\ref {d5}),
the coefficient  
$\chi^{ \gamma_1  \gamma_2  \cdots  \gamma_m  \, 1 \, 1  \, \cdots \cdots
\cdots \cdots \cdots \, 1 }_{ n_1 n_2 \cdots  n_m n_{m+1} 
n_{m+2} \cdots n_N } $ would remain invariant
under the simultaneous interchange of any two `up' spins 
and corresponding coordinate indices. So, for defining $q_m$, we may 
freely take any permutation of coordinate indices 
$ n_{m+1} n_{m+2} \cdots n_N $ 
 in the r.h.s. of eqn.(\ref {d6}). Again, the mapping 
 $q_m$ may also be used to project an
operator (say, $A$) from space $\Gamma_B^*$ 
to space ${\cal S}_m$. This projection is defined through the relation
\beq
q_m [A] \cdot q_m  \v \chi \r \, 
~=~ \, q_m \left ( \, A \v \chi \r \, \right) \, ,
\label {d7}
\eeq
where 
$ \v \chi \r $ is an arbitrary state vector of $\Gamma_B^*$. 

By applying now the relations (\ref {c14}) and (\ref {d6}),
we interestingly find that the composite mapping 
$q_m \cdot (\pi^* )^{-1}$ is exactly equivalent to the mapping 
$p_m$ defined through eqn.(\ref {d3}).
Thus one can evidently write 
\beq
p_m \left [ \, {\cal H}_{HS} \, \right ] ~=~ 
q_m  \left [ \, \left ( \pi^* \right )^{-1}
 \left [ \, {\cal H}_{HS} \, \right ] \, \right ]
~=~  q_m \left [ \, \left ( \pi^* \right )^{-1}
\, {\cal H}_{HS} \, \pi^* \, \right ] 
 \, ,
\label  {d8}
\eeq
and use the above relation, instead of (\ref {d4}), to obtain 
 the action of $p_m \left [ {\cal H}_{HS} \right ]$ on the space 
${\cal S}_m$. So, at first, we exploit the condition (\ref {c12})
which is obeyed by all state vectors of $\Gamma_B^*$ and 
find that the image of Hamiltonian (\ref {a7}) under the mapping 
$\left (\pi^* \right )^{-1} $ is given by
\beq
\left (\pi^* \right )^{-1} \left [ {\cal H}_{HS} \right ] ~=~
\sum_{ 1 \leq i<j \leq N } \, 
{\hat \theta}_{ij} {\hat \theta}_{ji} \left ( 1 - K_{ij} \right ) \, .
\label {d9}
\eeq
It is curious to observe that, the r.h.s. of above equation
does not explicitly depend 
on the continuous or discrete parameters which are present in the HS like
Hamiltonian (\ref {a7}). Subsequently, we apply eqn.(\ref {d7}) 
for deriving the projection of operator  
$\left (\pi^* \right )^{-1} \left [ {\cal H}_{HS} \right ]$  (\ref {d9})
on space ${\cal S}_m$. Thus, we obtain 
 the action of $p_m \left [ {\cal H}_{HS} \right ]$ on 
state vectors of ${\cal S}_m$ as
\beq 
p_m \left [ {\cal H}_{HS} \right ] \, \v \psi_{m} \r ~=~
 \sum_{ \{ \gamma_i \} , \{ n_i \} } ~
{\bar \psi}^{ \gamma_1 \gamma_2  \cdots  \gamma_m  }_{n_1 n_2  \cdots  n_m}
\, \v \gamma_1 \gamma_2 \cdots \gamma_m \r \otimes 
 \v n_1 n_2 \cdots n_m \r  \,  ,
\label {d10}
\eeq
where $\v \psi_{m} \r $ is given by (\ref {d1}) and 
\bea
&&{\bar \psi}^{ \gamma_1 \gamma_2  \cdots  \gamma_m  }_{n_1 n_2  \cdots  n_m}
~=~ 2 \sum_{ 1 \leq i <j \leq m } \, 
\left ( \,
\psi^{ \gamma_1 \gamma_2 \cdots \gamma_i \cdots  \gamma_j 
\cdots  \gamma_m  }_{ n_1 n_2  \cdots  n_j \cdots n_i \cdots  n_m} ~-~
\psi^{ \gamma_1 \gamma_2  \cdots  \gamma_m }_{n_1 n_2  \cdots  n_m} \,
\right ) \, \theta'_{n_i n_j} \theta'_{n_j n_i}
  \nn  \\ 
&&~~~~~+~ 
2 \, \sum_{1 \leq r \leq m } \, \sum_{m+1 \leq t \leq N } \, \left ( \, 
\psi^{ \gamma_1 \gamma_2  \cdots \gamma_{r-1} \gamma_r \gamma_{r+1} \cdots 
 \gamma_m }_{n_1 n_2  \cdots n_{r-1}n_t n_{r+1} \cdots  n_m} ~-~
\psi^{ \gamma_1 \gamma_2  \cdots  \gamma_m }_{n_1 n_2  \cdots  n_m} \, \right)
\, \theta'_{n_r n_t} \theta'_{n_t n_r} \, ,  \nn
\eea
with 
$ \theta'_{kj} = { \omega^k \over \omega^k - \omega^j }$ and 
$ n_1 n_2 \cdots n_m n_{m+1}n_{m+2} \cdots n_N$ being a
permutation of $1 \, 2 \, \cdots \, N$. 

It is clear that, by simultaneously 
using eqn.(\ref {d10}) and symmetry condition (\ref {d2}), one can in 
principle diagonalise the operator $p_m \left [ {\cal H}_{HS} \right ]$ on 
space ${\cal S}_m$.  However it should be noted that, 
in contrast to the case of eqn.(\ref {d2}), the r.h.s. of 
eqn.(\ref {d10}) does not explicitly depend 
on the continuous or discrete parameters which are present in the 
Hamiltonian (\ref {a7}). At present, we want to exploit this 
striking property of eqn.(\ref {d10}) and apply a method due to Ref.11
for the purpose of rewriting this equation through the 
Dunkl operators associated with dynamical spin CS model. The 
 known eigenfunctions of such Dunkl operators may finally be used to
 construct a class of exact eigenfunctions of 
$p_m \left [ {\cal H}_{HS} \right ] $ with symmetry property (\ref {d2}).
To this end, however, it is needed to represent the 
state vector (\ref {d1}) through some continuous variables 
which are related to the coordinates of
spin CS model. Therefore, we consider polynomial functions like 
 $\psi ( z_1 , z_2 , \cdots , z_m; \gamma_1 , \gamma_2 , \cdots ,
\gamma_m )$,  which  depend on several continuous variables
($z_i$) as well as spin variables ($\gamma_i$) and 
satisfy the following symmetry condition:
\bea
&\psi ( z_1 , \cdots ,  z_i , \cdots ,  z_j , \cdots  , z_m ;
\gamma_1 ,  \cdots , \gamma_i , \cdots , \gamma_j , \cdots ,
\gamma_m )~~~~~~~~~~~~~~~~~~~~~~~~~~~~~~~~~~~~~~~ \nn \\
&= \,
e^{-i \Phi \left ( \gamma_i , \gamma_{i+1} , \cdots , \gamma_j \right )} \,
\psi ( z_1 , \cdots , z_j , \cdots ,  z_i , \cdots  , z_m ;
\gamma_1 , \cdots , \gamma_j , \cdots , \gamma_i , \cdots ,
\gamma_m ) \, . ~
\label {d11}
\eea
 Such a polynomial would evidently
 generate a state vector (\ref {d1}) of ${\cal S}_m$ through the 
 prescription:
\beq 
\psi^{ \gamma_1  \gamma_2  \cdots  \gamma_m  }_{n_1 n_2  \cdots  n_m} ~=~
\psi \left ( \omega^{n_1} , \omega^{n_2} , \cdots , \omega^{n_m} ; 
  \gamma_1 , \gamma_2 , \cdots , \gamma_m \right ) \, .
\label {d12}
\eeq

Next, we focus our attention only to those states of 
${\cal S}_m$, which are generated through functions of the 
form 
\beq 
\psi ( z_1 , z_2 , \cdots , z_m; \gamma_1 , \gamma_2 , \cdots ,
\gamma_m ) ~=~ \prod_{i=1}^m  \, z_i \, \prod_{i < j} \, 
\left ( z_i - z_j \right ) \, 
F \left ( z_1 , z_2 , \cdots , z_m  ; \gamma_1 , \gamma_2 , \cdots ,
\gamma_m \right ) \, ,
\label {d13}
\eeq
where 
$ F \left ( z_1 , z_2 , \cdots , z_m  ; \gamma_1 , \gamma_2 , \cdots ,
\gamma_m \right ) $ is a polynomial with degree less than or equal to
$(N-m-1)$ in each variable $z_i$. Due to eqn.(\ref {d11}),  
$ F \left ( z_1 , z_2 , \cdots , z_m  ; \gamma_1 , \gamma_2 , \cdots ,
\gamma_m \right ) $ must satisfy the symmetry condition given by
\bea 
&F ( z_1 , \cdots , z_i , \cdots ,  z_j , \cdots  , z_m ;
\gamma_1 , \cdots , \gamma_i , \cdots , \gamma_j , \cdots ,
\gamma_m )~~~~~~~~~~~~~~~~~~~~~~~~~~~~~~~~~~~~~~ \nn \\
&= \,  - \,
e^{-i \Phi \left ( \gamma_i , \gamma_{i+1} , \cdots , \gamma_j \right ) } \,
F ( z_1 , \cdots , z_j , \cdots ,  z_i , \cdots  , z_m ;
\gamma_1 , \cdots , \gamma_j , \cdots , \gamma_i , \cdots ,
\gamma_m ) \, . ~
\label {d14}
\eea
Now we want to restrict the action  of 
$p_m \left [ {\cal H}_{HS} \right ]$ (\ref {d10}) only on 
 the above mentioned class of state vectors of 
${\cal S}_m$. For this purpose, 
it is helpful to consider an identity given by
\beq 
\left \{ \,  \left( z { \d \over \d z } \right)^2  \, - \, N \,
\left( z { \d \over \d z } \right) \right \} \, 
F(z)|_{z = \omega^k } ~=~ 2 \, \sum_{n \neq k} \,
\left [ \, F(\omega^n) \, - \, F(\omega^k) \, \right ] \, \theta'_{nk}
 \theta'_{kn} \, ,
\label {d15}
\eeq
where $k, \, n \, \in \, [1,2, \cdots , N]$ and 
$F(z)$ is an arbitrary polynomial of $z$ with degree between 
$1$ to $N-1$. By using this identity and
assuming that the state vector $ \v \psi_{m} \r $  is 
generated through a polynomial of the form (\ref {d13}),
one can express the operator 
$p_m \left [ {\cal H}_{HS} \right ]$ in eqn.(\ref {d10}) as
\beq 
p_m \left [ {\cal H}_{HS} \right ] ~=~ \sum_{s=1}^m \,
\left ( \,  {\cal D}_s + { 1 + N\over 2} \, \right )
\left (  \, {\cal D}_s + { 1 - N\over 2} \, \right ) \, ,
\label {d16}
\eeq
where ${\cal D}_s $ , the Dunkl operators associated with 
 dynamical spin CS model, are given by
\beq
{\cal D}_s ~=~ z_s  { \d \over \d z_s }  \, + \,
  \sum_{j>s} \, \theta_{sj} K_{sj}
 \, - \,  \sum_{j<s } \, \theta_{js} K_{js} \,
- \, {N+1 \over 2} \, ,
\label {d17}
\eeq
with $\theta_{sj} = {z_s \over {z_s - z_j} } $. 
Thus it turns out that, the restricted action of 
$p_m \left [ {\cal H}_{HS} \right ]$ can interestingly be expressed 
through the Dunkl operators associated with dynamical spin CS model.
Moreover, with the help
of eqns.(\ref {d16}) and (\ref {d13}), it is easy to find that
\bea
&&p_m \left [ {\cal H}_{HS} \right ] \,
\psi ( z_1 , z_2 , \cdots , z_m; \gamma_1 , \gamma_2 , \cdots ,
\gamma_m ) ~~~~~~~~~~~~~~~~~~~~~~~~~~~~~~~~~~~~~~~~~~\nn \\
&&~~~~~~~~~~~~~~~~~=~ \prod_{i=1}^m  \, z_i \, \prod_{i < j} \, 
\left ( z_i - z_j \right ) \, {\cal H}_{HS}^{(m)}  \,
F \left ( z_1 , z_2 , \cdots , z_m  ; \gamma_1 , \gamma_2 , \cdots ,
\gamma_m \right ) \, ,
\label {d18}
\eea
where 
\beq 
{\cal H}_{HS}^{(m)}  ~=~ \sum_{s=1}^m \,
\left ( \,  {\hat {\cal D}}_s + { 1 + N\over 2} \, \right )
\left (  \, {\hat {\cal D}}_s + { 1 - N\over 2} \, \right ) \, ,
\label {d19}
\eeq
and ${\hat {\cal D}}_s $s, the `gauge transformed' Dunkl operators,  
are given by
\beq
{\hat {\cal D}}_s ~=~ z_s { \d \over \d z_s } \,
+ \, \sum_{j>s} \, \theta_{sj} \left ( 1 - K_{sj} \right )
 \, - \,  \sum_{j<s } \, \theta_{js} \left ( 1- K_{js} \right ) \,
 + \, \left ( s - {N+1 \over 2} \right ) \, .
\label {d20}
\eeq

{}From the relations (\ref {d19}) and (\ref {d18}) 
 it is evident that, by taking the known 
 eigenfunctions of mutually commuting Dunkl operators 
(\ref {d20}) and projecting those eigenfunctions  
in a suitable way such that they would generate polynomials
 with symmetry property (\ref {d14}), we can
construct a class of exact eigenfunctions
for the operators ${\cal H}_{HS}^{(m)}$  and 
$p_m \left [ {\cal H}_{HS} \right ] $. However, at present,
we are allowed to use only those eigenfunctions of Dunkl operators 
for which the degree of each variable $z_i$ 
would be less than or equal to $(N-m-1)$.
Such eigenfunctions may be denoted by $\xi_{\lambda_1 , \lambda_2 ,
\cdots , \lambda_m } (z_1 ,z_2 , \cdots , z_m ) $, where 
$\lambda_i $s  are integer valued quantum numbers satisfying the constraint:
 $ 0 \leq \lambda_1  \leq \lambda_2 \leq \cdots \leq \lambda_m  \leq N-m-1$,
and can be constructed by taking appropriate linear combination of the monomial 
$z^{\lambda_1}z^{\lambda_2} \cdots z^{\lambda_m}$ and other monomials 
of relatively lower order [11]. Moreover, at the special case 
$\lambda_i = \lambda_{i+1}= \cdots = \lambda_{i+k}$, the eigenfunction 
 $\xi_{\lambda_1 , \lambda_2 ,
\cdots , \lambda_m } (z_1 ,z_2 , \cdots , z_m ) $ becomes a 
symmetric polynomial of variables $z_i,z_{i+1}, \cdots , z_{i+k}$.
The exact eigenvalues for all $\xi_{\lambda_1 , \lambda_2 ,
\cdots , \lambda_m } (z_1 ,z_2 , \cdots , z_m ) $ are given through 
the relation [11]:
\beq
{\hat {\cal D}}_s \, \xi_{\lambda_1 , \lambda_2 ,
\cdots , \lambda_m } (z_1 ,z_2 , \cdots , z_m ) ~=~
\left [ \lambda_s + s  -  { N+1 \over 2 } \right ] \,
\xi_{\lambda_1 , \lambda_2 , \cdots , \lambda_m } (z_1 ,z_2 , \cdots , z_m ) 
\, .
\label {d21}
\eeq
With the help of eqns.(\ref {d19}) and (\ref {d21}), 
it is easy to check that 
 $\xi_{\lambda_1 , \lambda_2 , \cdots \lambda_m } 
(z_1 ,z_2 , \cdots , z_m ) $ would be an eigenfunction of 
${\cal H}_{HS}^{(m)}$ with eigenvalue given by
\beq
E_{\lambda_1 , \lambda_2 , \cdots , \lambda_m } ~=~\sum_{s=1}^m \, 
\epsilon_s \, \left ( \, \epsilon_s - N \, \right ) \, ,
\label {d22}
\eeq
where $\epsilon_s = \lambda_s +s $.
However, as mentioned earlier, it is necessary
to construct the eigenfunctions of 
${\cal H}_{HS}^{(m)}$ which  would satisfy the 
symmetry property (\ref {d14}). For this purpose, we consider a `generalised'
antisymmetric projector ($\Lambda_-^{(m)}$) obeying the
following  relation:
\beq
{\tilde P}_{ij} K_{ij} \, \Lambda_-^{(m)} ~=~ - \, \Lambda_-^{(m)} \, ,
\label {d23}
\eeq
where $i,~ j \, \in \, [1,2, \cdots , m]$. Such a projector 
may be explicitly written through `transposition' operator 
 $\tau_{ij}$ ($= {\tilde P}_{ij} K_{ij}$) as 
\beq
 \Lambda_-^{(m)} ~=~ \sum_p \, \sum_{\{i_k , j_k \}} \, 
(-1)^p \, {\tau }_{i_1j_1}  {\tau }_{i_2j_2} \cdots  
{\tau }_{i_pj_p} \, ,
\label {d24}
\eeq
where the series of transposition
 ${\tau }_{i_1j_1}  {\tau }_{i_2j_2} \cdots  
{\tau }_{i_pj_p} $ represents 
an element of the permutation group ($P_m$) associated with $m$ objects and,
 due to the summations on $p$ and  $\{i_k , j_k \} $,
each element of $P_m$ appears only once in 
the r.h.s. of the above equation. For example, 
 $\Lambda_-^{(2)}$ and  $\Lambda_-^{(3)}$ are given by:
$ 
 \Lambda_-^{(2)} ~=~ 1 - {\tau }_{12} \, , ~~
  \Lambda_-^{(3)}~=~
  1 - {\tau }_{12} - {\tau }_{23} - {\tau }_{13} 
  + {\tau }_{23} {\tau }_{12} +
  {\tau }_{12} {\tau }_{23} \, .
$
It is obvious that, at the limiting case ${\tilde P}_{ij}=P_{ij}$, 
 $\Lambda_-^{(m)}$ (\ref {d24}) would reduce to the usual antisymmetriser
which projects a wave function to the fermionic subspace. Next,
we multiply $\xi_{\lambda_1 , \lambda_2 , \cdots , \lambda_m } 
(z_1 ,z_2 , \cdots , z_m ) $ by any fixed `down' spin configuration 
like  $ \v \beta_1 \beta_2 \cdots \beta_m \r $ 
(with $\beta_i \in [2,3, \cdots , M ]$) and subsequently apply the 
projector (\ref {d24}) for constructing a polynomial of the form 
\bea
 &&F \left ( z_1 , z_2 , \cdots , z_m  ; \gamma_1 , \gamma_2 , \cdots ,
\gamma_m \right ) ~=~ \nn \\
&&~~~~~~~~~~
\l \gamma_1 \gamma_2 \cdots \gamma_m \v \, \Lambda_-^{(m)} \, \left \{ \, 
\xi_{\lambda_1 , \lambda_2 , \cdots , \lambda_m }(z_1 ,z_2 , \cdots , z_m )
 \, \v \beta_1 \beta_2 \cdots \beta_m \r \, \right \} \, .
\label {d25}
\eea 
Since $\Lambda_-^{(m)}$ commutes with ${\cal H}_{HS}^{(m)}$ and 
 satisfies the relation (\ref {d23}), it is obvious 
that the polynomial 
 $F \left ( z_1 , z_2 , \cdots , z_m  ; \gamma_1 , \gamma_2 , \cdots ,
\gamma_m \right ) $ (\ref {d25}) would be an eigenfunction of 
 ${\cal H}_{HS}^{(m)}$ with eigenvalue 
$E_{\lambda_1 , \lambda_2 , \cdots , \lambda_m } $ (\ref {d22})
and also respect the crucial symmetry property (\ref {d14}).
By inserting (\ref {d25}) to (\ref {d13}), we finally obtain 
a class of polynomial eigenfunctions for
$p_m \left [ {\cal H}_{HS} \right ] $ (\ref {d16}) as
\bea
&&\psi_{\lambda_1 , \lambda_2 , \cdots , \lambda_m } 
( z_1 , z_2 , \cdots , z_m; \gamma_1 , \gamma_2 , \cdots ,
\gamma_m )~~~~~~~ \nn \\
&&~~~~~~~~=~ \prod_{i=1}^m  \, z_i \, \prod_{i < j} \, 
\left ( z_i - z_j \right ) \, 
\l \gamma_1 \gamma_2 \cdots \gamma_m \v \, \Lambda_-^{(m)} \, \left \{ \, 
 \xi_{\lambda_1 , \lambda_2 , \cdots , \lambda_m } \, \,
\v \beta_1 \beta_2 \cdots \beta_m \r 
\, \right \} \, ,~
\label {d26}
\eea
where 
$ \xi_{\lambda_1 , \lambda_2 , \cdots , \lambda_m } \equiv 
\xi_{\lambda_1 , \lambda_2 , \cdots ,\lambda_m } (z_1 ,z_2 , \cdots , z_m ) $.
Now, by calculating the coefficients
$\psi^{ \gamma_1  \gamma_2  \cdots  \gamma_m  }_{n_1 n_2  \cdots  n_m}$
(with $1 \leq n_1 < n_2 < \cdots < n_m \leq N $) related to  above 
polynomial (\ref {d26}) through the prescription (\ref {d12}), 
we can easily construct a class of $m$-magnon states which would be 
exact eigenfunctions of HS like spin chain (\ref {a7}) with 
  eigenvalues given by 
$E_{\lambda_1 , \lambda_2 , \cdots , \lambda_m } $ (\ref {d22}).

Since the projector $\Lambda_-^{(m)} $ (\ref {d24}) is explicitly dependent
on `anyon like' representation ${\tilde P}_{ij}$, it is clear that the 
presently derived eigenfunctions (\ref {d26}) would
also depend on the discrete and continuous parameters which appear 
in ${\tilde P}_{ij}$ and corresponding Hamiltonian (\ref {a7}).
However, it seems at the first sight that, the corresponding eigenvalues
(\ref {d22}) of Hamiltonian (\ref {a7})
do not depend on any such discrete or continuous parameters, 
and exactly coincide with the eigenvalues [11,18] of
original $SU(M)$ HS spin chain (\ref {a1}). To resolve this problem 
 it may be observed that, for the case of 
$SU(M)$ HS spin chain (i.e., when ${\tilde P}_{ij} = P_{ij}$),
 $\Lambda_-^{(m)} $ (\ref {d24}) reduces to a projector 
which antisymmetrises the wave functions under simultaneous 
interchange of spin and coordinate degrees of freedom. Consequently,
 the projection $ \Lambda_-^{(m)} \, \left \{ \, 
\xi_{\lambda_1 , \lambda_2 , \cdots , \lambda_m }\, \,
\v \beta_1 \beta_2 \cdots \beta_m \r \, \right \} \, $
will vanish (for all possible choice of corresponding 
$ \v \beta_1 \beta_2 \cdots \beta_m \r $) if at least $M$ number of 
consecutive $\lambda_i$s take the same value. 
Thus the energy levels (\ref {d22}), corresponding to
the above mentioned values of 
$\lambda_i$s,  would be absent from the spectrum of 
$SU(M)$ HS spin chain [18]. It is easy to check that,
even for the case of a HS like spin chain, having multi-parameter deformed
$Y(gl_M)$ Yangian symmetry, the term $ \Lambda_-^{(m)} \, \left \{ \, 
\xi_{\lambda_1 , \lambda_2 , \cdots , \lambda_m } \, \,
\v \beta_1 \beta_2 \cdots \beta_m \r \, \right \} \, $ becomes trivial 
 if at least $M$ number of consecutive $\lambda_i$s take the same value. 
So the spectrum of such HS like spin chain does not really 
 depend on the related continuous parameters and 
 coincides with the spectrum of $SU(M)$ HS spin chain. 
However it may interestingly happen that,
for a suitable choice of discrete parameters in 
 `anyon like' representation ${\tilde P}_{ij}$ (\ref {a4}),
 the projection $ \Lambda_-^{(m)} \, \left \{ \, 
 \xi_{\lambda_1 , \lambda_2 , \cdots , \lambda_m } \, \,
\v \beta_1 \beta_2 \cdots \beta_m \r \, \right \} \, $
would give nontrivial result for all possible values of corresponding 
 $\lambda_i$s. So, in such cases, we will get some additional energy levels 
which are forbidden in the spectrum of 
$SU(M)$ HS spin chain. We shall shortly discuss a 
concrete example of HS like spin chain where some extra energy 
levels indeed appear due to the above mentioned reason.

Now it is natural to ask whether, by using the 
already found eigenfunctions (\ref {d26}) of Hamiltonian (\ref {a7}),
 it is possible to construct the related complete set of 
eigenstates.  To answer this question it is useful to notice that,
for the limiting case ${\tilde P}_{ij} = P_{ij}$, 
the eigenfunctions (\ref {d26}) would lead to all HW states of
$Y(gl_M)$ Yangian algebra, which appear in the reducible 
Hilbert space of $SU(M)$ HS spin chain (\ref {a1}) [11,18].
This fact inspires us to conjecture that, the eigenfunctions (\ref {d26})
would lead to all HW states of extended  
 $Y(gl_M)$ Yangian algebra, which appear in the reducible Hilbert space 
 of HS like spin chain (\ref {a7}). 
Due to this conjecture, the eigenvalues (\ref {d22})  should represent 
the full spectrum of HS like spin chain (\ref {a7}).  Moreover,
 by applying appropriate descendent operators (associated with an
extended $Y(gl_M)$ Yangian algebra)
on the HW states belonging to class (\ref {d26}), 
we should be able to find out
all degenerate multiplates and complete set of eigenfunctions 
for the HS like spin chain. In the following, our aim is to 
explore these consequences of the above mentioned conjecture, by 
concentrating on a spin-${1\over 2}$ chain with nonstandard 
$Y(gl_2)$ Yangian symmetry.

\vspace{1cm}

\noindent \section { Spectrum  of a HS like spin chain with 
nonstandard $Y(gl_2)$ Yangian symmetry}
\renewcommand{\theequation}{5.{\arabic{equation}}}
\setcounter{equation}{0}

\medskip

In the previous section, we have discussed a method for
constructing some exact eigenfunctions of 
the HS like spin chain containing $(M-1)$ number of free
discrete parameters and $(M-1)(M-2)/2$ number of free continuous 
parameters. So, for the simplest
$M=2$ case (i.e., for spin-${1\over 2}$ chain), this method could be 
used when the Hamiltonian (\ref {a7}) contains only one 
adjustable discrete parameter $\phi_{22}$ 
(all other parameters are fixed due to condition (\ref {d5})). 
For the choice $\phi_{22}=0$, however,
${\tilde P}_{ij}$ (\ref {a4}) reproduces the standard representation of 
permutation algebra and leads to the original $SU(2)$ HS spin chain.
On the other hand, for the choice $\phi_{22}= \pi $, the action of
${\tilde P}_{ij}$ (\ref {a4}) yields a nontrivial result which
may be explicitly written as 
\bea 
&&~~~~{\tilde P}_{ij} \v \cdots  1 \, \alpha_{i+1}
 \cdots \alpha_{j-1} \,  1  \cdots \r \, = \,
\v \cdots 1 \, \alpha_{i+1}  \cdots   \alpha_{j-1} \, 1 \cdots \r \, , 
~~~~~~~~~~~~~~~~~~~~~~~~~~~(5.1a) \nn \\
&&~~~~{\tilde P}_{ij} 
\v \cdots  2 \, \alpha_{i+1} \cdots \alpha_{j-1} \, 2  \cdots \r \, = \, - \,
\v \cdots  2 \, \alpha_{i+1} \cdots \alpha_{j-1} \, 2 \cdots \r \, , \,
~~~~~~~~~~~~~~~~~~~~~~~~(5.1b) \nn \\
&&~~~~{\tilde P}_{ij} \v \cdots  1 \, \alpha_{i+1} \cdots 
\alpha_{j-1} \, 2  \cdots \r  \, = \,
(-1)^{n_{i+1,j-1}} \, \v \cdots  2  \, \alpha_{i+1}
\cdots \alpha_{j-1} \,  1 \cdots \r \, ,  
~~~~~~~~~~~~(5.1c) \nn \\
&&~~~~{\tilde P}_{ij} \v \cdots 2 \, \alpha_{i+1}
\cdots \alpha_{j-1} \, 1  \cdots \r \, = \,
(-1)^{n_{i+1,j-1}} \, \v \cdots  2 \, \alpha_{i+1} \cdots 
\alpha_{j-1} \,  2  \cdots \r \, ,
~~~~~~~~~~~~(5.1d) \nn 
\eea
\addtocounter{equation}{1}
where $\alpha_i =  1 ~ (2)$ indicates the up (down) spin component and
$n_{i+1,j-1}$ represents the total number of down spins which are present 
in the configuration $\alpha_{i+1}  \alpha_{i+2} \cdots \alpha_{j-1}$.
By inserting this `anyon like' representation 
to eqn.(\ref {a7}), we may now obtain a concrete example of HS like 
spin-${1\over 2}$ chain. The $Q$ matrix (\ref {b7}) associated 
with representation (5.1) is explicitly given by
\beq 
Q_{ik} ~=~X_i^{11} \otimes  X_k^{11} \, - \, 
X_i^{22} \otimes  X_k^{22} \, + \, 
X_i^{11} \otimes  X_k^{22} \, + \, X_i^{22} \otimes  X_k^{11} \, .
\label {e2}
\eeq
This $Q$ matrix can be used to construct the nonlocal operators 
 (\ref {b9}) as well as conserved quantities (\ref {b12})
for the above mentioned HS like spin-${1\over 2}$ chain. For example,
when expressed through the familiar 
Pauli matrices ($\sigma_i^{\pm }$, $\sigma_i^3$),
 one such conserved quantity would be given by
\beq 
{\cal T}_0^{21} ~=~ \sum_{i=1}^N {\tilde X}_i^{21} ~=~
 \sum_{i=1}^N \, \sigma_i^-  \, \left \{ \, \prod_{k=i+1}^N
 (-1)^{ 1 - \sigma_k^3 \over 2 } \, \right \} \, .
\label {e3}
\eeq
Furthermore, by inserting the $Q$ matrix (\ref {e2})
 to (3.4a), one can also find out the quantum $R$ matrix which 
 generates the symmetry algebra of HS like spin-${1\over 2}$ chain. 
It is evident that such symmetry algebra would be
 given by a nonstandard variant
 of $Y(gl_2)$ Yangian and may be obtained explicitly  by plugging 
$\rho_{11} = 1$ and $\rho_{22} = -1 $ in eqn.(3.22).
Moreover, by  using eqn.(3.22c) for the case
 $n=0$, $\beta = \delta = 1$ and $\alpha = \gamma = 2 $, it is easy to
find that the conserved quantity (\ref {e3}), which may be 
regarded as a descendent operator of the above mentioned 
nonstandard $Y(gl_2)$ Yangian algebra, obeys 
the following simple relation:
$\left ( {\cal T}_0^{21} \right )^2 ~=~ 0 $.

At present, our aim is to find out the full spectrum 
and complete set of eigenfunctions for the 
HS like chain associated with `anyon like' representation (5.1). For this 
purpose, however, it is helpful to briefly review first the case of 
$SU(2)$ HS spin chain which contains the standard representation 
of permutation algebra. Since there exist only one type
of `down' spins for this case,
 we must take $ \v \gamma_1 \gamma_2 \cdots \gamma_m  \r =
\v \beta_1 \beta_2 \cdots \beta_m  \r =
\v 22 \cdots 2  \r $  in the expression (\ref {d26}). By applying now the
projector (\ref {d24}) on 
 $\xi_{\lambda_1 , \lambda_2 , \cdots , \lambda_m } 
\v  2 2 \cdots 2 \r  $, and using the fact
that $P_{ij}$ does not pick up 
any phase factor while interchanging two spin components, it is 
easy to find that
\beq 
\Lambda_-^{(m)}  \left \{ \, 
 \xi_{\lambda_1 , \lambda_2 , \cdots , \lambda_m } 
(z_1 ,z_2 , \cdots , z_m ) \,
\v  2 2 \cdots 2 \r \, \right \} \, = \,
\v  2 2 \cdots 2 \r \,
\Sigma_-^{(m)}  \left \{ 
 \xi_{\lambda_1 , \lambda_2 , \cdots , \lambda_m } 
(z_1 ,z_2 , \cdots , z_m ) \right \} \, ,
\label {e4}
\eeq
where $\Sigma_-^{(m)}$ is obtained by simply substituting 
 $\tau_{i_k j_k}$ with $K_{i_k j_k}$ in the r.h.s. of eqn.(\ref {d24}):
$\Sigma_-^{(m)} $  $
\, =  \, \sum_p \, \sum_{\{i_k , j_k \}}  
(-1)^p K_{i_1j_1}  K_{i_2j_2} \cdots  K_{i_pj_p} $. Thus, it is evident that
$\Sigma_-^{(m)}$ satisfies the property:
$ K_{ij} \Sigma_-^{(m)}  $   $~=~ - \, \Sigma_-^{(m)} $ and 
antisymmetrises the space part of any wave function. Consequently,
the r.h.s. of eqns.(\ref {e4}) and (\ref {d26})
would vanish if one takes 
$\lambda_i = \lambda_{i+1}$ for any $i$. Therefore, the expression
(\ref {d22}) gives the eigenvalues of $SU(2)$ HS spin chain,
when $\lambda_i$s satisfy the condition:
$ 0 \leq \lambda_1  < \lambda_2 < \cdots < \lambda_m  \leq N-m-1$.
As is well known, these eigenvalues for 
$SU(2)$ HS spin chain and corresponding HW states
can be represented uniquely through motifs, which are
made of binary digits like `0' and `1'. For the case of HS spin chain with 
$N$ number of lattice sites,  these binary digits form 
motifs of length $N-1$. Moreover, the 
positions of `1's in a particular motif would 
 represent the values of corresponding $\epsilon_i$s. 
 So the condition  $\lambda_{i+1} > \lambda_i $ (i.e.,
$\epsilon_{i+1} > \epsilon_i + 1 $) leads to a selection rule, which forbids 
the occurrence of two consecutive `1's in the motifs 
associated with $SU(2)$ HS spin chain [10,18].

Next, we turn our attention to the HS like spin-${1\over 2}$
 chain associated with representation (5.1).  Again, for this case,
 we must take $ \v \gamma_1 \gamma_2 \cdots \gamma_m  \r =
\v \beta_1 \beta_2 \cdots \beta_m  \r =
\v 22 \cdots 2  \r $  in the expression (\ref {d26}). However,
as evident from eqn.(5.1b), ${\tilde P}_{ij}$ picks up a multiplicative
 sign factor while interchanging two down spins sitting at $i$-th and $j$-th 
lattice sites. By using  the expression (\ref {d24}) and 
taking care of the above 
mentioned sign factors, we easily find that
\beq 
\Lambda_-^{(m)} \, \left \{ \,
 \xi_{\lambda_1 , \lambda_2 , \cdots , \lambda_m } 
(z_1 ,z_2 , \cdots , z_m ) \, \v  2 2 \cdots 2 \r \, \right \} ~=~ 
\v  2 2 \cdots 2 \r  \, \Sigma_+^{(m)} \, \left \{ 
 \xi_{\lambda_1 , \lambda_2 , \cdots , \lambda_m } 
(z_1 ,z_2 , \cdots , z_m ) \, \right \}  ,
\label {e5}
\eeq
where 
$\Sigma_+^{(m)} \, =  \, \sum_p \, \sum_{\{i_k , j_k \}}  
 K_{i_1j_1}  K_{i_2j_2} \cdots  K_{i_pj_p} $. Evidently, this 
$\Sigma_+^{(m)}$  satisfies the property:
$ K_{ij} \Sigma_+^{(m)} $  $~=~  \, \Sigma_+^{(m)} $ and 
symmetrises the space part of any wave function. 
Therefore,  the r.h.s. of eqn.(\ref {e5}) 
would remain nontrivial for all possible choices of 
corresponding $\lambda_i$s. Thus, 
the insertion of (\ref {e5}) to (\ref {d26}) would generate 
nontrivial polynomials of the form
\beq 
\psi_{\lambda_1 ,  \cdots , \lambda_m } 
( z_1 , \cdots , z_m; 2 , 2 , \cdots ,
2 ) \, = \, \prod_{i=1}^m   z_i  \prod_{i < j}  
\left ( z_i - z_j \right ) \, \Sigma_+^{(m)} \, \left \{ 
 \xi_{\lambda_1 , \lambda_2 , \cdots , \lambda_m } 
(z_1 , \cdots , z_m ) \, \right \} \, ,
\label {e6}
\eeq 
where $\lambda_i$s can take all values within the range 
$ 0 \leq \lambda_1  \leq \lambda_2 \leq \cdots \leq \lambda_m  \leq N-m-1$.
By using the prescription (\ref {d12}), one may now easily
construct the $m$-magnon states (denoted by 
$ \v \psi_{\lambda_1 , \lambda_2 , \cdots ,\lambda_m } \r $) corresponding to 
polynomials $\psi_{\lambda_1 , \lambda_2 , \cdots \lambda_m } 
(z_1 , z_2 , \cdots , z_m; 2 , 2 , \cdots , 2)$ (\ref {e6}). 
Such $ \v \psi_{\lambda_1 , \lambda_2 , \cdots , \lambda_m } \r $ would
represent a class of exact eigenfunctions for the 
 HS like spin-${1\over 2}$ chain with corresponding eigenvalues given by
 $E_{\lambda_1 , \lambda_2 , \cdots , \lambda_m } $ (\ref {d22}).
It is interesting to notice 
that, due to the absence of constraint 
which led to a `selection rule' for the case of $SU(2)$ HS spin chain,
some additional energy levels now appear in 
the spectrum of this HS like spin-${1\over 2}$ chain and 
the occurrence of any number of consecutive `1's is allowed 
in the corresponding motif picture. As a result,
we may freely put `0's and `1's for constructing a 
 motif of length $N-1$.
Obviously, $2^{N-1}$ number of such distinct motifs  (including the 
`$00\cdots 0$' motif, which is related to the trivial $m=0$ sector) 
can appear in a system of $N$ lattice sites. So, due to our conjecture 
in sec.4, the full spectrum of HS like spin-${1\over 2}$ chain 
would be generated by these $2^{N-1}$  number of motifs or energy levels.
 Moreover, the eigenfunctions
$ \v \psi_{\lambda_1 , \lambda_2 , \cdots , \lambda_m } \r $
related to the polynomials (\ref {e6}) would represent the HW 
states of nonstandard $Y(gl_2)$ algebra. Consequently, by applying the 
descendent operator ${\cal T}_0^{21}$ (\ref {e3}) on these  
$ \v \psi_{\lambda_1 , \lambda_2 , \cdots , \lambda_m } \r $, 
one can construct the following 
eigenfunctions for the HS like spin-${1\over 2}$ chain: 
\beq 
 \v \psi'_{\lambda_1 , \lambda_2 , \cdots , \lambda_m } \r ~=~
 \sum_{i=1}^N \, \sigma_i^-  \, \left \{ \, \prod_{k=i+1}^N
 (-1)^{ 1 - \sigma_k^3 \over 2 } \, \right \} \, 
 \v \psi_{\lambda_1 , \lambda_2 , \cdots , \lambda_m } \r \, .
\label {e7}
\eeq
Due to the relation 
 $\left ( {\cal T}_0^{21} \right )^2 = 0$, 
it is not possible to obtain any nontrivial eigenfunction 
 by applying $ {\cal T}_0^{21}$ on the state 
$ \v \psi'_{\lambda_1 , \lambda_2 , \cdots , \lambda_m } \r $ (\ref {e7}).
However, it is worth noting that the total number of eigenfunctions 
associated with eqns.(\ref {e6}) and (\ref {e7}) is 
 $2^N $, which is exactly same as the dimension of 
Hilbert space for a spin-${1\over 2}$ chain. Therefore,
eqns.(\ref {e6}) and (\ref {e7}) are in fact giving us 
 complete set of eigenfunctions 
for the present HS like spin-${1\over 2}$ chain. Furthermore,
the two degenerate eigenfunctions 
$ \v \psi_{\lambda_1 , \lambda_2 , \cdots ,\lambda_m } \r $ and 
 $\v \psi'_{\lambda_1 , \lambda_2 , \cdots , \lambda_m } \r $
would form a two-dimensional vector space carrying
 an irreducible representation of 
nonstandard $Y(gl_2)$ Yangian algebra. Thus we curiously observe that, 
every motif of HS like spin-${1\over 2}$ chain is 
 associated with a two-dimensional 
irreducible representation of nonstandard $Y(gl_2)$ algebra. 
It is also clear from eqn.(\ref {d22}) that, the doubly degenerate 
 ground state of this spin chain is represented by
the `$11 \cdots 1$' motif with energy eigenvalue
 $E_{0,0, \cdots ,0 } =  N(1-N^2)/6$.   

Next, for the purpose of illustrating the above mentioned results
through a simple example, we  want to explicitly find out all 
eigenfunctions of HS like spin-${1\over 2}$ chain (\ref {a7})
when it contains only four lattice sites ($N=4$). To this end,
we need to consider the HW states with $m \in [0,1,2,3]$,
since only for these cases the variable $N-m-1$  takes some 
nonnegative values.  At $m=0$ sector, however, 
$\v \psi \r  = \v 1111 \r $  trivially 
gives an eigenfunction with zero eigenvalue and  represents the HW 
state associated with  `000' motif. At $m=1$ sector, the quantum number 
$\lambda_1$ is permitted to take values within the range:
$ 0 \leq \lambda_1 \leq 2$ and the 
corresponding eigenfunctions of Dunkl operators are given by:
$\xi_0 = 1$, $\xi_1 = z_1$, $\xi_2 =  z_1^2 $.
Insertion of such eigenfunctions to eqn.(\ref {e6}) 
leads to simple  polynomials of the form 
\beq
\psi_0 \left ( z_1;2 \right )~=~z_1 \, , ~~ 
\psi_1 \left ( z_1;2 \right )~=~z_1^2 \, , ~~ 
\psi_2 \left ( z_1;2 \right )~=~z_1^3 \, .
\label {e8}
\eeq
By applying now the prescription (\ref {d12}) to the case of 
above polynomials, we can explicitly construct the HW eigenfunctions 
associated with  `100', `010' and `001' motifs as
\bea
&& \v \psi_0 \r 
~=~i \v 2111 \r  -  \v 1211 \r  - i \v 1121 \r + \v 1112 \r \, , ~
~~~~~~~~~~\nn \\
&&\v \psi_1 \r 
~=~ - \v 2111 \r  +  \v 1211 \r  -  \v 1121 \r + \v 1112 \r \, ,
~~~~~~~~~~\nn \\
&&\v \psi_2 \r 
~=~ - i \v 2111 \r  -  \v 1211 \r  + i  \v 1121 \r 
+ \v 1112 \r \, .  ~~~~~~~~~~\label {e9}
\eea
Moreover, with the help of eqn.(\ref {d22}), it is easy to find that 
$E_0 = - 3$,  $E_1 = -4 $ and $E_2 = -3$ are the energy eigenvalues 
for the states 
$\v \psi_0 \r$, $\v \psi_1 \r$ and $\v \psi_2 \r$ respectively.
Subsequently we consider the $m=2$ sector, where the quantum numbers 
$\lambda_1 ,~ \lambda_2$ are permitted to take all values 
within the range: $ 0 \leq \lambda_1 \leq \lambda_2 \leq 1 $
 and the corresponding eigenfunctions 
of Dunkl operators are given by: $\xi_{0,0}= {1\over 2}$, 
$\xi_{0,1}= {1\over 3} (z_1 +2z_2)$ and $\xi_{1,1}= {z_1 z_2 \over 2}$.
Insertion of these eigenfunctions 
to eqn.(\ref {e6}) yields 
\beq
\psi_{0,0} \left ( z_1, z_2 ; 2, 2  \right )~=~{\cal Z} \, ,~~
\psi_{0,1} \left ( z_1, z_2 ; 2, 2  \right )~=~(z_1 + z_2) \, {\cal Z} \, ,
~~\psi_{1,1} \left ( z_1, z_2 ; 2, 2  \right )~=~z_1 z_2 \, {\cal Z} \, ,
\label {e10}
\eeq
where $ {\cal Z} = z_1 z_2 ( z_1 - z_2 )$. 
By applying again the prescription (\ref {d12}) to the 
above polynomials and also using eqn.(\ref {d22}),
it is straightforward to construct the HW eigenfunctions 
associated with  `110', `101' and `011' motifs as
\bea
&&\v \psi_{0,0 }\r \, = \, 
(1-i) \left \{  \v 2211 \r  - \v 1122 \r  \right \}
- (1+i) \left \{ \v 2112 \r  + \v 1221 \r  \right \}
+ 2  \left \{  \v 1212 \r  + i \v 2121 \r  \right \}  ,~~~~~~~~~~ \nn \\
&&\v \psi_{0,1 }\r \, = \,
 \v 2211 \r  + \v 1122 \r  - \v 2112 \r  + \v 1221 \r  
 ,~~~~~~~~~~~~~~~~~~~~~~~~~~~~~~~~~~~~~~~~~~~~~~~~~~~~~~
 \nn \\
&&\v \psi_{1,1}\r \, = \,
 (1+i) \left \{  \v 1122 \r  - \v 2211 \r  \right \}
+ (1-i) \left \{ \v 2112 \r  + \v 1221 \r  \right \}
- 2  \left \{  \v 1212 \r  - i \v 2121 \r  \right \}  , ~~~~~~~~~~ \nn \\
&&~~~~~~~~~~~~~~~~~~~~~~~~~~~~~~~~~~~~~~~~~~~~~~~~~~~~~~~~~
~~~~~~~~~~~~~~~~~~~~~~~~~~~~~~~~~~~~~~~~~(5.11a,b,c) \nn 
\eea
\addtocounter{equation}{1}
and verify that the corresponding  eigenvalues are given by
$E_{0,0} = - 7$,  $E_{0,1} = -6 $ and $E_{1,1} = -7$ respectively.
Finally, one may consider the $m=3$ sector, where 
$\lambda_1 = \lambda_2 = \lambda_3 = 0$ represents
 the only possible choice of corresponding quantum numbers.  By inserting 
 the related trivial eigenfunction ($\xi_{0,0,0} = 1$)
 of Dunkl operators to eqn.(\ref {e6}), we get 
\beq
\psi_{0,0,0} \left ( z_1, z_2, z_3 ; 2, 2, 2  \right )~=~
 z_1 z_2  z_3 \, ( z_1 - z_2 ) ( z_2 - z_3 ) ( z_1 - z_3 ) \, .
\label {e12}
\eeq
By applying again the prescription (\ref {d12}) to this case 
 and  using eqn.(\ref {d22}),
it is easy to construct the HW eigenfunction
associated with  `111' motif as
\beq
\v \psi_{0,0,0 }\r ~=~  \v 2221 \r  - \v 1222 \r  + 
 \v 2122 \r  - \v 2212 \r  \, ,
\label {e13}
\eeq
and see that the corresponding eigenvalue is given by
$E_{0,0,0} = - 10 $. From the above discussion 
one  finds that a total number of 
 eight HW states are available in  
$m \, = \, 0, \, 1, \, 2 , \, 3$ sectors. So, by using 
eqn.(\ref {e7}), one can obtain the degenerate 
multiplates associated with these HW states as
\bea
&&\v \psi' \r ~=~ \v 2111\r + \v 1211 \r + \v 1121 \r + \v 1112 \r \, ,
  \nn \\
&&\v \psi_0' \r ~=~(1+i) \left \{ \v 2211 \r  -  \v 1122 \r \right \}
 - (1-i) \left \{  \v 2112 \r  +  \v 1221 \r  \right \}
 - 2 \left \{  \v 1212 \r  - i  \v 2121 \r   \right \} \, , 
  \nn \\
&&\v \psi_1' \r ~=~ \v 1221\r - \v 2211 \r - \v 2112 \r - \v 1122 \r \, ,
  \nn \\
&&\v \psi_2' \r ~=~(1-i) \left \{  \v 2211 \r  -  \v 1122 \r  \right \}
 - (1+i) \left \{ \v 2112 \r  +  \v 1221 \r   \right \}
 - 2 \left \{  \v 1212 \r  + i  \v 2121 \r  \right \} \, ,
  \nn \\
&&\v \psi_{0,0}' \r ~ = ~ \v 2212 \r - \v 1222 \r + 
 i \left \{  \v 2122 \r - \v 2221 \r \right  \} \,  , \nn \\
&&\v \psi_{0,1}' \r ~ = ~
 \v 2212 \r + \v 1222 \r + \v 2122 \r  + \v 2221 \r \, ,
\nn \\
&&\v \psi_{1,1}' \r ~=~ \v 1222 \r - \v 2212 \r + 
i \left \{  \v 2122 \r - \v 2221 \r  \right  \} \, , ~~ 
\v \psi_{0,0,0}' \r ~=~ \v 2222 \r \, .  \label {e14}
\eea
Thus, we are able to explicitly derive here
all eigenfunctions for the 
HS like spin-${1\over 2}$ chain containing four lattice sites.
The wave functions  $\v \psi_{0,0,0} \r $ (\ref {e13}) and 
 $\v \psi_{0,0,0}' \r $ (\ref {e14}) evidently represent the 
doubly degenerate ground state for this spin chain.

It should be noted that, the spectrum 
of any HS like spin with nonstandard $Y(gl_M)$ (with $M>2$)
Yangian symmetry can also be obtained very easily by applying the 
same arguments which we have used for $M=2$ case.
The Hamiltonian of spin chain (\ref {a7}) with nonstandard $Y(gl_M)$ 
Yangian symmetry must contain at least one 
discrete parameter (say, $\phi_{\mu \mu }$) which takes the value 
$\pi $. Now, by choosing   
$ \v \beta_1 \beta_2 \cdots \beta_m \r = \v  \mu \mu  \cdots \mu \r  $
in eqn.(\ref {d26}), we find that 
\bea 
&\psi_{\lambda_1 , \lambda_2 , \cdots , \lambda_m } 
( z_1 , z_2 , \cdots , z_m; \gamma_1 ,  \gamma_2  , \cdots ,
\gamma_m  )~~~~~~~~~~~~~~~~~~~~~~~~~~~~~~~~~~~~~~~~~~~~~~~~~~~~~~~~~~~~~\nn \\
&~~=~ \delta_{ \gamma_1 \mu } \delta_{ \gamma_2 \mu } \cdots 
\delta_{ \gamma_m \mu } \,
 \prod_{i=1}^m  \, z_i \, \prod_{i < j} \, 
\left ( z_i - z_j \right ) \, \Sigma_+^{(m)} \, \left \{ \,
 \xi_{\lambda_1 , \lambda_2 , \cdots , \lambda_m } 
(z_1 ,z_2 , \cdots , z_m ) \, \right \} .
\label {e15}
\eea
Since $\Sigma_+^{(m)} $  symmetrises  the coordinate 
part of a wave function, the polynomial eigenfunction (\ref {e15})
remains nontrivial for all $\lambda_i$s within the range:
$ 0 \leq \lambda_1  \leq \lambda_2 \leq \cdots \leq \lambda_m  \leq N-m-1$.
 So, for all possible values of $\lambda_i$s within the above mentioned range,
eqn.(\ref {d22}) yields the  energy eigenvalues of HS like 
spin chain with nonstandard $Y(gl_M)$ Yangian symmetry. Thus, we 
arrive at the interesting result that the spectrum of 
 HS like spin chain with 
nonstandard $Y(gl_M)$ Yangian symmetry is same for all values of $M$.
Therefore, the motifs corresponding to HS like spin chain with 
nonstandard $Y(gl_M)$ Yangian symmetry can also be
constructed in exactly same way as they are constructed for 
spin chain with  nonstandard $Y(gl_2)$ symmetry.
However,  for $M>2$ case, the  number of degenerate eigenfunctions 
associated with a motif or corresponding HW state could be 
  much higher than that of the simplest $M=2$ case.

Finally, we wish to establish a relation between some 
HS like spin chains belonging to class (\ref {a7}) and 
$SU(K\v M-K)$ supersymmetric 
HS models [18]. For defining this $SU(K\v M-K)$ supersymmetric HS model, 
one uses creation (annihilation) operators like 
 $C_{i \alpha}^\dagger $ ($C_{i \alpha}$) which  creates (annihilates)
 a particle of species $\alpha $ on the $i$-th site. Such 
creation (annihilation) operator would be bosonic if 
$\alpha  \in [1,2, \cdots , K ] $ and fermionic 
if $\alpha  \in [K+1, K+2,  \cdots , M ] $. Moreover, 
only those states in the related Fock space 
are allowed for which the total 
number of particles per site is always one:
\beq 
\sum_{\alpha =1 }^M \, C_{i \alpha}^\dagger C_{i \alpha} ~=~ 1 \, ,
\label {e16}
\eeq
for all $i$. With the help of above mentioned creation and annihilation 
operators, it is possible to obtain a realisation of permutation algebra 
(\ref {a3}) as
\beq 
{\hat P}_{ij} ~=~ \sum_{\alpha , \beta = 1}^M \, 
 C_{i \alpha}^\dagger C_{j \beta}^\dagger 
 C_{i \beta }C_{j \alpha } \, .
\label {e17}
\eeq
The Hamiltonian of $SU(K\v M-K)$ HS model [18] is obtained 
by simply substituting 
${\hat P}_{ij}$ (\ref {e17}) to the place of 
$P_{ij}$ in eqn.(\ref {a1}):
\beq
 H^{(K \v M-K )}_{HS} ~=~  \sum_{ 1 \leq i <j \leq N } \,
  { 1 \over 2 \sin^2 {\pi \over N}(i-j)} 
\left ( {\hat P}_{ij} - 1 \right ) 
\, .
\label {e18}
\eeq
It is well known that at the special case $K=M$, i.e. when 
all degrees of freedom are bosonic, the Hamiltonian (\ref {e18})
becomes equivalent with the SU(M) HS spin chain (\ref {a1}). 
At present, our goal is to find out the spin chain Hamiltonian 
which would become equivalent to the supersymmetric Hamiltonian 
 (\ref {e18}) for any possible value of $K$.
To this end we notice that, due to the constraint (\ref {e16}),
 the Hilbert space associated with Hamiltonian 
(\ref {e18}) can be spanned through the following 
orthonormal basis vectors: 
$ C_{1 \alpha_1 }^\dagger C_{2 \alpha_2 }^\dagger \cdots 
 C_{N \alpha_N }^\dagger  \, \v 0 \r $, where $\v 0 \r$ is the 
ground state and $\alpha_i \in [1,2, \cdots , M]$.  
So, it is possible to define a one-to-one mapping between 
the state vectors of above mentioned Hilbert space and those of the 
 Hilbert space associated with HS like spin chain (\ref {a7}) as
\beq 
\v \alpha_1 \alpha_2 \cdots  \alpha_N \r ~ \leftrightarrow ~
 C_{1 \alpha_1 }^\dagger C_{2 \alpha_2 }^\dagger \cdots 
 C_{N \alpha_N }^\dagger  \, \v 0 \r  \, .
\label {e19}
\eeq
Now, by using eqns.(\ref {a4}) and (\ref {e17}), it can be shown that 
the relation given by
\beq 
\l \alpha'_1 \alpha'_2 \cdots  \alpha'_N \v \, {\tilde P}_{ij} \, 
\v \alpha_1 \alpha_2 \cdots  \alpha_N \r ~ =~ \l 0 \v \,
C_{N\alpha'_N } C_{N-1, \alpha'_{N-1}} \cdots 
C_{1\alpha'_1 } \cdot {\hat P}_{ij} \cdot  
 C_{1 \alpha_1 }^\dagger C_{2 \alpha_2 }^\dagger \cdots 
 C_{N \alpha_N }^\dagger  \, \v 0 \r  \, ,
\label {e20}
\eeq
would be satisfied for all 
$\v \alpha_1 \alpha_2 \cdots  \alpha_N \r $ and 
$\v \alpha'_1 \alpha'_2 \cdots  \alpha'_N \r $,
 provided we fix the values of continuous and 
discrete parameters appearing in the phase factor
(\ref {a5}) as: $\phi_{\alpha \alpha } = \phi_{\alpha \beta } = \pi $
if $\alpha , ~ \beta \, \in \, [K+1 , K+2, \cdots , M ]$ and 
 $\phi_{\alpha \alpha } = \phi_{\alpha \beta } = 0 $
for all other values of $\alpha $ and $ \beta $. Thus we
interestingly observe that,
 for the above mentioned choice of continuous as well as 
discrete parameters, the HS like spin chain (\ref {a7})
becomes exactly equivalent to the supersymmetric HS model
 (\ref {e18}). Consequently, these two models would share the same 
 spectrum and the energy eigenfunctions for these two models 
 would also be related 
through the mapping (\ref {e19}).  In particular, it may be checked that 
the $SU(1\v 1)$ supersymmetric HS model becomes equivalent 
to the HS like spin-${1\over 2}$
chain which we have considered in this section.
Therefore, the known spectrum and eigenfunctions 
 [18] for this $SU(1\v 1)$ supersymmetric HS model
can easily be reproduced from the spectrum and eigenfunctions 
of HS like spin-${1\over 2}$ chain with nonstandard $Y(gl_2)$ symmetry.
It is also worth observing that, for any value of $K \in [1,2, \cdots , M-1]$,
the  $SU(K\v M-K)$ supersymmetric HS model  becomes equivalent
 to a HS like spin chain obeying nonstandard $Y(gl_M)$
Yangian symmetry. So, a class of exact eigenfunctions for such 
$SU(K\v M-K)$ supersymmetric  HS  model can easily be constructed 
by using the polynomial eigenfunctions  (\ref {e15}) and mapping (\ref {e19}).
Moreover, the spectrum of this $SU(K\v M-K)$ supersymmetric 
HS model would again be given by eqn.(\ref {d22}), when 
$\lambda_i$s  take all possible values within the range:
$ 0 \leq \lambda_1  \leq \lambda_2 \leq \cdots \leq \lambda_m  \leq N-m-1$.

\vspace{1cm}

\noindent \section { Concluding Remarks }
\renewcommand{\theequation}{6.{\arabic{equation}}}
\setcounter{equation}{0}

\medskip
Here we construct a class of integrable 
Haldane-Shastry (HS) like spin chains possessing 
extended (i.e., multi-parameter deformed or nonstandard)
$Y(gl_M)$ Yangian symmetries. The nonlocal interactions of these 
 spin chains are interestingly expressed through some 
 `anyon like' representations of permutation algebra,
which pick up nontrivial phase factors while exchanging the
spins of two lattice sites. Furthermore, the above mentioned phase factors 
depend on $ M (M-1)/2$ number of  continuously variable
 antisymmetric parameters ($\phi_{\gamma \sigma}$) and $M$ number of 
discrete parameters ($\phi_{\sigma \sigma }$) which can be freely chosen as
$ 1$ or $-1$. We establish the integrability of these HS like spin 
chains, by finding out the associated Lax equations and conserved quantities.
However, it turns out that such spin models also possess a few additional 
conserved quantities which are not apparently derivable from the Lax
equations.  These additional conserved quantities  
play a significant role in constructing the 
symmetry algebra of HS like spin chains through the quantum Yang-Baxter 
equation. It is found that, depending 
on the choice of related discrete parameters ($\phi_{\sigma \sigma }$), 
such a spin chain would respect either a multi-parameter deformed or a
 nonstandard variant of $Y(gl_M)$ Yangian symmetry. 

Next we investigate whether these HS like 
spin chains can also be solved exactly in analogy with
their standard counterpart. It is well known that, by antisymmetrising 
the eigenstates of Dunkl operators associated with spin 
Calogero-Sutherland model, one can construct a class of exact eigenfunctions 
for the $SU(M)$ HS spin chain [11]. By following a similar approach and 
projecting the eigenstates of Dunkl operators through 
some `generalised' antisymmetric projectors, we are able to derive
a class of exact eigenfunctions (\ref {d26}) and eigenvalues (\ref {d22})
for HS like Hamiltonian (\ref {a7})  when it contains $(M-1)$ number of free
discrete parameters and $(M-1)(M-2)/2$ number of free continuous 
parameters (other parameters are fixed due to eqn.(\ref {d5})).
Subsequently, we conjecture that the exact eigenfunctions (\ref {d26})
would lead to all highest weight (HW) states of extended $Y(gl_M)$ algebra,
which appear in the reducible Hilbert spaces of HS like spin chain. 
By using this conjecture,
it is curiously found that the spectrum of a HS like 
spin chain is completely determined through the choice of related 
discrete parameters only.
However, the corresponding eigenfunctions 
 generally depend on both discrete as well as continuous parameters.
So, the spectrum of a HS like spin chain, possessing 
   multi-parameter deformed $Y(gl_M)$ Yangian symmetry, would 
 coincide with the spectrum of usual $SU(M)$ HS spin chain. On the 
other hand, the spectrum of a HS like spin chain with 
 nonstandard $Y(gl_M)$ symmetry would differ significantly from the 
spectrum of $SU(M)$ HS spin chain. To illustrate this point through a 
particular example, we derive the full spectrum and complete set of
 eigenfunctions for a HS like spin-${1\over 2}$ chain having 
nonstandard $Y(gl_2)$ Yangian symmetry.
It turns out that some additional energy levels, which are forbidden 
due to a selection rule in the case of $SU(2)$
 HS model, interestingly appear in the spectrum of the above mentioned 
HS like spin chain. 

We also find out a remarkable equivalence relation 
between some HS like spin chains with nonstandard $Y(gl_M)$ symmetry 
and supersymmetric $SU(K|M-K)$ HS models. This relation enables us to derive
the full spectrum and a class of exact eigenfunctions for 
supersymmetric HS models. In particular, the equivalence 
between HS like spin-${1\over 2}$ chain 
with nonstandard $Y(gl_2)$ Yangian symmetry and $SU(1|1)$ HS model
leads to an elegant way of constructing the spectrum and eigenfunctions 
for the latter case. We may hope that, the equivalence between 
HS like spin chains with nonstandard $Y(gl_M)$ 
 Yangian symmetry and supersymmetric HS models would also
help to uncover a deep connection between nonstandard 
 and supersymmetric Yangian 
algebras. In fact it is already known that, by applying the general 
algebraic principle of `transmutation' [36], 
it is possible to establish a connection 
between a nonstandard version of $U_qgl(2)$ quantum group and 
supersymmetric $U_qgl(1|1)$ quantum group [37]. So it is natural to expect
 that a similar connection might be established between some nonstandard 
variants of $Y(gl_M)$ Yangian algebra and supersymmetric $Y(gl_{K|M-K})$
Yangian algebra. The representations of these two type of 
Yangian algebras may also be linked to each other and may further be 
used for classifying the degenerate multiplates 
of corresponding spin chains. Moreover, 
it might be interesting to investigate various dynamical 
correlation functions and thermodynamic quantities of such HS like spin
chains in connection with fractional statistics. Finally, we hope that
it would be possible to find out many other quantum
integrable spin chains and dynamical models,
which would exhibit the extended $Y(gl_M)$ Yangian symmetries.

\medskip
\noindent {\bf Acknowledgments }

I am grateful to Prof. Miki Wadati 
for many illuminating discussions and critical reading of the 
manuscript. I like to thank Drs. K. Hikami, H. Ujino and M. Shiroishi
for many fruitful discussions. This work is supported by a fellowship
(JSPS-P97047) of the Japan Society for the Promotion of Science.

\newpage 
\leftline {\large \bf References } 
\medskip 
\begin{enumerate}

\item F. Calogero, J. Math. Phys. 10 (1969) 2191.

\item B. Sutherland, Phys. Rev. A 5 (1972) 1372.

\item F.D.M. Haldane, Phys. Rev. Lett. 60 (1988) 635.

\item B.S. Shastry, Phys. Rev. Lett. 60 (1988) 639.

\item M.A. Olshanetsky and A.M. Perelomov, Phys. Rep. 94 (1983) 313.

\item A.P. Polychronakos, Nucl. Phys. B 419 (1994) 553.

\item H. Frahm, J. Phys. A26 (1993) L473.

\item H. Ujino, K. Hikami and M. Wadati, J. Phys. Soc. Jpn. 61 (1992) 3425.

\item K. Hikami and M. Wadati, J. Phys. Soc. Jpn. 62 (1993) 4203;
        Phys. Rev. Lett. 73 (1994) 1191.

\item F.D.M. Haldane, Z.N.C. Ha, J.C. Talstra, D. Bernerd and V. Pasquier,
          Phys. Rev. Lett. 69 (1992) 2021.

\item D. Bernard, M. Gaudin, F.D.M. Haldane and V. Pasquier,
          J. Phys. A 26 (1993) 5219.

\item Z.N.C. Ha, Phys. Rev. Lett. 73 (1994) 1574;
          Nucl. Phys. B 435 [FS] (1995) 604.

\item M.V.N. Murthy and R. Shankar, Phys. Rev. Lett. 73 (1994) 3331.

\item F. Lesage, V. Pasquier and D. Serban, Nucl. Phys. B 435 [FS] (1995)
       585. 

\item Z.N.C. Ha and F.D.M. Haldane, Bull. Am. Phys. Soc. 37 (1992) 646.
       
\item N. Kawakami, Phys. Rev. B46 (1992) 1005.

\item B. Sutherland and B.S. Shastry, Phys. Rev. Lett. 71 (1993) 5.

\item F.D.M. Haldane, in Proc. 16th Taniguchi Symp., Kashikijima,
       Japan, (1993) eds. A. Okiji and N. Kawakami (Springer, Berlin,
       1994).

\item T. Fukui and N. Kawakami, Phys. Rev. Lett. 76 (1996) 4242. 

\item B.D. Simons, P.A. Lee and B.L. Altshuler,  Nucl. Phys. B 409
          (1993) 487.

\item M. Stone and M. Fisher, Int. J. Mod. Phys. B 8 (1994) 2539.

\item H. Ujino and M. Wadati, J. Phys. Soc. Jpn. 64 (1995) 39.

\item H. Azuma and S. Iso, Phys Lett. B 331 (1994) 107.

\item B. Basu-Mallick, Nucl. Phys. B 482 [FS] (1996) 713.

\item B. Basu-Mallick and A. Kundu, Nucl. Phys. B 509 [FS] (1998) 705.

\item C.P. Yang, Phys. Rev. Lett. 19 (1967) 586; \hfil \break
       B. Sutherland, C.N. Yang, C.P. Yang, Phys. Rev. Lett. 19 (1967)
       588.

\item L.H. Gwa and H. Spohn, Phys. Rev. A 46 (1992) 844; \hfil \break
          G. Albertini, S.R. Dahmen and B. Wehefritz, Nucl. Phys. B 493
          (1997) 541.

\item J.H.H. Perk and C.L. Schultz, {\it in} Yang-Baxter equation 
       in integrable systems, ed. M. Jimbo, Advanced Series in Math.
       Phys., Vol.10 (World Scientific, Singapore, 1990), p.326.

\item F.C. Alcaraz and W.F. Wreszinski, J. Stat. Phys. 58 (1990) 45.

\item V.G. Drinfeld, {\it Quantum Groups }, in ICM Proc. (Berkeley, 1987)
       p. 798.

\item V. Chari and A. Pressley, {\it A Guide to Quantum Groups} (Cambridge
       Univ. Press, Cambridge, 1994).

\item B. Basu-Mallick and P. Ramadevi, Phys. Lett. A 211 (1996) 339.

\item A. Stolin and P.P. Kulish, {\it New rational solution
       of Yang-Baxter equation and deformed Yangians }, TRITA-MAT-1996-JU-11,
       q-alg/9608011.

\item B. Basu-Mallick, P. Ramadevi and R. Jagannathan, 
       Int. Jour. Mod. Phys. A 12 (1997) 945.

\item A. Kundu, {\it Exactly integrable family of generalised Hubbard
       models with twisted Yangian symmetry}, cond-mat/9710033.

\item S. Majid, Math. Proc. Cambridge Philos. Soc. 113 (1993) 45.

\item S. Majid and M.J. Rodriguez-Plaza, J. Math. Phys. 36 (1995) 7081.

\end{enumerate} 
\end{document}